\documentclass{ws-mpla}

\newcommand{\re}{\mathop{\mathrm{Re}}\nolimits}

\begin{document}

\markboth{Mathias Butenschoen, Bernd A. Kniehl}
{Next-to-leading-order tests of nonrelativistic-QCD factorization with $J/\psi$
yield and polarization}

\catchline{}{}{}{}{}

\title{
\vskip-4cm{\baselineskip14pt
\centerline{\normalsize DESY 12--235\hfill ISSN 0418-9833}
\centerline{\normalsize December 2012\hfill}}
\vskip2.5cm
\boldmath NEXT-TO-LEADING-ORDER TESTS OF NONRELATIVISTIC-QCD FACTORIZATION WITH
$J/\psi$ YIELD AND POLARIZATION\unboldmath}

\author{\footnotesize MATHIAS BUTENSCHOEN}

\address{Universit\"at Wien, Fakult\"at f\"ur Physik,\\ Boltzmanngasse 5,
1090 Wien, Austria\\
mathias.butenschoen@univie.ac.at}

\author{BERND A. KNIEHL}

\address{II. Institut f\"ur Theoretische Physik, Universit\"at Hamburg,\\
Luruper Chaussee 149, 22761 Hamburg, Germany\\
kniehl@desy.de}

\maketitle

\pub{Received (Day Month Year)}{Revised (Day Month Year)}

\begin{abstract}
We report on recent progress in testing the factorization formalism
of nonrelativistic quantum chromodynamics (NRQCD) at next-to-leading order
(NLO) for $J/\psi$ yield and polarization.
We demonstrate that it is possible to unambiguously determine the leading
color-octet long-distance matrix elements (LDMEs) in compliance with the
velocity scaling rules through a global fit to experimental data of unpolarized
$J/\psi$ production in $pp$, $p\overline{p}$, $ep$, $\gamma\gamma$, and
$e^+e^-$ collisions.
Three data sets not included in the fit, from hadroproduction and from
photoproduction in the fixed-target and colliding-beam modes, are nicely
reproduced.
The polarization observables measured in different frames at DESY HERA and CERN
LHC reasonably agree with NLO NRQCD predictions obtained using the LDMEs
extracted from the global fit, while measurements at the FNAL Tevatron exhibit
severe disagreement.
We demonstrate that alternative LDME sets recently obtained, with different
philosophies, in two other NLO NRQCD analyses of $J/\psi$ yield and polarization
also fail to reconcile the Tevatron polarization data with the other available
world data.

\end{abstract}

\ccode{PACS Nos.: 12.38.Bx, 13.60.Le, 13.88.+e, 14.40.Pq}

\section{Introduction}

While the overly successful experiments at the LHC are exploring the Higgs
sector and are systematically searching for signals of physics beyond the
standard model (SM), we must not be carried away losing track of a
longstanding, unresolved puzzle in quantum chromodynamics (QCD), the otherwise
well-established SU(3) gauge theory of the strong interactions, right in the
core of the SM.
In fact, despite concerted experimental and theoretical efforts ever since the
discovery of the $J/\psi$ meson in the November revolution of 1974 (The Nobel
Prize in Physics 1976), the genuine mechanism underlying the production and
decay of heavy quarkonia, which are QCD bound states of a heavy quark $Q=c,b$
and its antiparticle $\overline{Q}$, has remained mysterious.

Nonrelativistic QCD (NRQCD)\cite{Caswell:1985ui} endowed with an appropriate
factorization theorem, which was conjectured in a seminal work by Bodwin,
Braaten, and Lepage\cite{Bodwin:1994jh} and explicitly proven through
next-to-next-to-leading order for large transverse momenta
$p_T$,\cite{Nayak:2005rt,Nayak:2006fm} arguably constitutes the most probable
candidate theory at the present time.
This implies a separation of process-dependent short-distance coefficients, to
be calculated perturbatively as expansions in the strong-coupling constant
$\alpha_s$, from supposedly universal long-distance matrix elements
(LDMEs), to be extracted from experiment.
The relative importance of the latter can be estimated by means of velocity
scaling rules,\cite{Lepage:1992tx} which predict each of the LDMEs to scale
with a definite power of the heavy-quark velocity $v$ in the limit $v\ll1$.
In this way, the theoretical predictions are organized as double expansions in
$\alpha_s$ and $v$.
A crucial feature of this formalism is that the $Q\overline{Q}$ pair can at
short distances be produced in any Fock state
$n={}^{2S+1}L_J^{[a]}$ with definite spin $S$, orbital angular momentum
$L$, total angular momentum $J$, and color multiplicity $a=1,8$.
In particular, this formalism predicts the existence of intermediate
color-octet (CO) states in nature, which subsequently evolve into physical,
color-singlet (CS) quarkonia by the nonperturbative emission of soft gluons.
In the limit $v\to0$, the traditional CS model (CSM) is recovered in the case
of $S$-wave quarkonia.
In the case of $J/\psi$ production, the CSM prediction is based just on the
$^3\!S_1^{[1]}$ CS state, while the leading relativistic corrections, of
relative order ${\cal O}(v^4)$, are built up by the $^1\!S_0^{[8]}$,
$^3\!S_1^{[8]}$, and $^3\!P_J^{[8]}$ ($J=0,1,2$) CO states.

The CSM is not a complete theory, as may be understood by noticing that the
next-to-leading-order (NLO)
treatment of $P$-wave quarkonia is plagued by uncanceled infrared
singularities, which are, however, properly removed in NRQCD.
This conceptual problem cannot be cured from within the CSM, neither by
proceeding to higher orders nor by invoking $k_T$ factorization {\it etc.}
In a way, NRQCD factorization,\cite{Bodwin:1994jh} appropriately improved at
large values of $p_T$ by systematic expansion in powers of
$m_Q^2/p_T^2$,\cite{Kang:2011zz,Kang:2011mg,Fleming:2012wy} is the only game
in town, which makes its experimental verification such a matter of paramount
importance and general interest.\cite{Brambilla:2010cs}

The experimental test of NRQCD factorization\cite{Bodwin:1994jh} has been
among the most urgent tasks on the agenda of the international quarkonium
community\cite{Brambilla:2010cs} for almost two decades and, with
high-quality data being so copiously harvested at the LHC, is now more
tantalizing than ever.
In the following, we discuss the present status of testing NRQCD factorization
in charmonium production.

\section{\boldmath Global fit to measurements of unpolarized $J/\psi$ yields
\unboldmath}

\begin{table}[ht]
\tbl{NLO NRQCD fit results for the $J/\psi$ CO
LDMEs.\protect\cite{Butenschoen:2011yh}
Subtracting from the data the estimated contributions from the feed-down of
heavier charmonia, which are not included in the calculations, improves the
quality of the fit.}
{\begin{tabular}{@{}ccc@{}} \toprule
 & set A: unsubtracted & set B: subtracted \\
\colrule
$\langle {\cal O}^{J/\psi}(^1\!S_0^{[8]}) \rangle$ &
$(4.97\pm0.44)\times10^{-2}$~GeV$^3$ &
$(3.04\pm0.35)\times10^{-2}$~GeV$^3$ \\
$\langle {\cal O}^{J/\psi}(^3\!S_1^{[8]}) \rangle$ &
$(2.24\pm0.59)\times10^{-3}$~GeV$^3$ &
$(1.68\pm0.46)\times10^{-3}$~GeV$^3$ \\
$\langle {\cal O}^{J/\psi}(^3\!P_0^{[8]}) \rangle$ &
$(-1.61\pm0.20)\times10^{-2}$~GeV$^5$ &
$(-9.08\pm1.61)\times10^{-3}$~GeV$^5$ \\
$\chi_{\rm d.o.f.}^2$ & 4.42 & 3.74 \\
\botrule
\end{tabular}\label{tab:fit}}
\end{table}

We consider the inclusive production of $J/\psi$ mesons in collisions of two
particles $A$ and $B$.
Owing to the factorization theorems of the QCD parton model and
NRQCD,\cite{Bodwin:1994jh} the cross section is calculated as
\begin{eqnarray}
d\sigma(AB\to J/\psi+X)
&=&\sum_{i,j,k,l,n}\int dx_1dx_2dy_1dy_2\,f_{i/A}(x_1)f_{k/i}(y_1)
f_{j/B}(x_2)f_{l/j}(y_2)
\nonumber\\
&&{}\times\langle{\cal O}^{J/\psi}[n]\rangle
d\sigma(k l\to c\overline{c}[n]+X),
\label{eq:xs}
\end{eqnarray}
where $f_{i/A}(x_1)$ is the parton distribution function (PDF) of parton
$i=g,q,\overline{q}$ in hadron $A=p,\overline{p}$ or the flux function of
photon $i=\gamma$ in charged lepton $A=e^-,e^+$, $f_{k/i}(y_1)$ is
$\delta_{ik}\delta(1-y_1)$ or the PDF of parton $k$ in the resolved photon $i$,
$d\sigma(k l\to c\overline{c}[n]+X)$ are the partonic cross sections, and
$\langle{\cal O}^{J/\psi}[n]\rangle$ are the LDMEs.
In the fixed-flavor-number scheme, we have $q=u,d,s$.
In the case of $e^+e^-$ annihilation, all distribution functions in
Eq.~(\ref{eq:xs}) are delta functions.
The hadronic system $X$ always contains one hard parton at leading order (LO)
and is taken to be void of heavy flavors, which may be tagged and vetoed
experimentally.\cite{Pakhlov:2009nj,Aaij:2012dz}
The partonic cross sections appropriate for the direct production of
unpolarized $J/\psi$ mesons were calculated at NLO in NRQCD in
Refs.~\refcite{Butenschoen:2009zy,Butenschon:2010iy} for direct
photoproduction, in
Refs.~\refcite{Butenschoen:2010rq,Butenschoen:2010px,Ma:2010yw,Ma:2010jj} for
hadroproduction, and in Ref.~\refcite{Butenschoen:2011yh} for resolved
photoproduction, two-photon scattering involving both direct and resolved
photons, and $e^+e^-$ annihilation. 

In our numerical analysis,
we set $m_c=1.5$~GeV, adopt the values of $m_e$, $\alpha$, and the branching
ratios  $B(J/\psi\to e^+e^-)$ and $B(J/\psi\to\mu^+\mu^-)$ from
Ref.~\refcite{Nakamura:2010zzi}, and use the one-loop (two-loop) formula for
$\alpha_s^{(n_f)}(\mu)$, with $n_f=4$ active quark flavors, at LO (NLO).
As for the proton PDFs, we use set CTEQ6L1 (CTEQ6M)\cite{Pumplin:2002vw} at
LO (NLO), which comes with an asymptotic scale parameter of
$\Lambda_\mathrm{QCD}^{(4)}=215$~MeV (326~MeV).
As for the photon PDFs, we employ the best-fit set AFG04\_BF of
Ref.~\refcite{Aurenche:2005da}.
We evaluate the photon flux function using Eq.~(5) of
Ref.~\refcite{Kniehl:1996we}, with the upper cutoff on the photon virtuality $Q^2$
chosen as in the considered data set.
As for the CS LDME, we adopt the value
$\langle {\cal O}^{J/\psi}(^3\!S_1^{[1]}) \rangle = 1.32$~GeV$^3$ from
Ref.~\refcite{Bodwin:2007fz}.
Our default choices for the renormalization, factorization, and NRQCD scales
are $\mu_r=\mu_f=m_T$ and $\mu_\Lambda=m_c$, respectively, where
$m_T=\sqrt{p_T^2+4m_c^2}$ is the $J/\psi$ transverse mass.
The bulk of the theoretical uncertainty is due to the lack of knowledge of
corrections beyond NLO, which are estimated by varying $\mu_r$, $\mu_f$, and
$\mu_\Lambda$ by a factor 2 up and down relative to their default values.

In Ref.~\refcite{Butenschoen:2011yh}, we performed a global fit to high-quality
data of inclusive unpolarized $J/\psi$ production, comprising a total of 194
data points from 26 data sets.
Specifically, these included $p_T$ distributions in hadroproduction from
PHENIX\cite{Adare:2009js} at RHIC, CDF at
Tevatron~I\cite{Abe:1997jz,Abe:1997yz} and II,\cite{Acosta:2004yw}
ATLAS,\cite{ATLASdata,Kirk} CMS,\cite{Khachatryan:2010yr}
ALICE,\cite{Scomparin:2011zzb} and LHCb\cite{Aaij:2011jh} at the LHC; $p_T^2$,
$W$, and $z$ distributions in photoproduction from H1\cite{Adloff:2002ex} and
ZEUS\cite{Chekanov:2002at} at HERA~I and H1\cite{Aaron:2010gz} at HERA~II; a
$p_T^2$ distribution in two-photon scattering from DELPHI\cite{Abdallah:2003du}
at LEP~II; and a total cross section in $e^+e^-$ annihilation from
Belle\cite{Pakhlov:2009nj} at KEKB.
Denoting the photon, proton, and $J/\psi$ four-momenta by $p_\gamma$, $p_p$,
and $p_{J/\psi}$, respectively, $W=\sqrt{(p_\gamma+p_p)^2}$ is the $\gamma p$
center-of-mass energy and $z=(p_{J/\psi}\cdot p_p)/(p_\gamma\cdot p_p)$ is the
inelasticity variable measuring the fraction of the photon energy passed on to
the $J/\psi$ meson in the proton rest frame.
We excluded from our fit all data points of two-photon scattering with
$p_T<1$~GeV and of hadroproduction with $p_T<3$~GeV, which cannot be
successfully described by our fixed-order calculations as expected.
The fit results for the CO LDMEs obtained at NLO in NRQCD with default scale
choices are collected in Table~\ref{tab:fit}.
They depend only feebly on the precise locations of the $p_T$ cuts.

Our calculations refer to direct $J/\psi$ production, as the data from
Tevatron~I\cite{Abe:1997jz,Abe:1997yz} do, while the data from
Tevatron~II,\cite{Acosta:2004yw}
LHC,\cite{ATLASdata,Kirk,Khachatryan:2010yr,Scomparin:2011zzb,Aaij:2011jh} and
KEKB\cite{Pakhlov:2009nj} comprise prompt events and those from
RHIC,\cite{Adare:2009js} HERA,\cite{Adloff:2002ex,Chekanov:2002at,Aaron:2010gz}
and LEP~II\cite{Abdallah:2003du} even non-prompt ones.
The fit results obtained neglecting the effects due to these admixtures are
listed in the second column of Table~\ref{tab:fit} (set A).
However, the resulting error is small against our theoretical uncertainties and
has no effect on our conclusions.
In fact, the fraction of $J/\psi$ events originating from the feed-down of
heavier charmonia only amounts to about 36\% for
hadroproduction,\cite{Abe:1997jz} 15\% for photoproduction at
HERA,\cite{Aaron:2010gz} 9\% for two-photon scattering at
LEP~II,\cite{Klasen:2004tz} and 26\% for $e^+e^-$ annihilation at
KEKB,\cite{Ma:2008gq} and the fraction of $J/\psi$ events from $B$ decays is
negligible RHIC, HERA,\cite{Aaron:2010gz} and LEP~II\cite{Klasen:2004tz}
energies.
Refitting the data with the estimated feed-down contributions subtracted yields
the values listed in the third column of Table~\ref{tab:fit} (set B).
The $\chi^2$ values per data point achieved by the two fits, which are
specified as $\chi_{\rm d.o.f.}^2$ in Table~\ref{tab:fit}, are to be taken with
a grain of salt, since they do not take into account the theoretical
uncertainties, which exceed most of the experimental errors.

The fact that the global fit\cite{Butenschoen:2011yh} successfully pins down
the three CO LDMEs as it does is quite nontrivial by itself and establishes
their universality, the more so as the long-standing difficulty of NRQCD to
describe the photoproduction data at large values of $z$ is overcome.
Furthermore, their values are of order ${\cal O}(v^4)$ with respect to the CS
LDME $\langle{\cal O}^{J/\psi}(^3\!S_1^{[1]})\rangle$,\cite{Bodwin:2007fz} 
in compliance with the velocity scaling rules.\cite{Lepage:1992tx}
Both observations consolidate the validity of NRQCD factorization as far as the
unpolarized $J/\psi$ yield is concerned.

\begin{figure}[ph]
\begin{center}
\includegraphics[width=\textwidth]{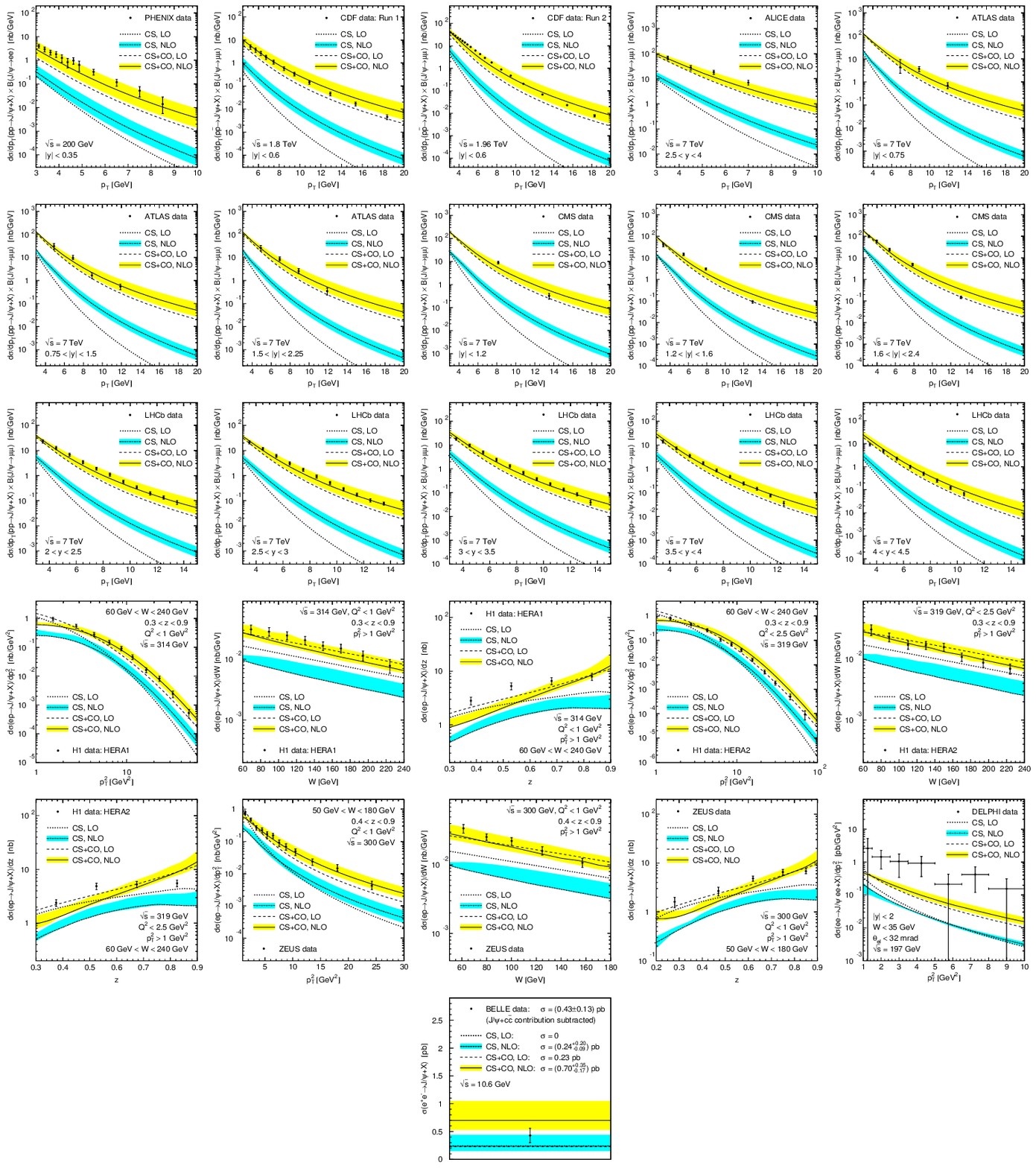}
\end{center}
\vspace*{8pt}
\caption{NLO NRQCD fit\protect\cite{Butenschoen:2011yh} compared to
RHIC,\protect\cite{Adare:2009js}
Tevatron,\protect\cite{Abe:1997jz,Abe:1997yz,Acosta:2004yw}
LHC,\protect\cite{ATLASdata,Kirk,Khachatryan:2010yr,Scomparin:2011zzb,Aaij:2011jh}
HERA,\protect\cite{Adloff:2002ex,Chekanov:2002at,Aaron:2010gz}
LEP~II,\protect\cite{Abdallah:2003du}, and KEKB\protect\cite{Pakhlov:2009nj}
data.
\protect\label{fig:fit}}
\end{figure}

In Fig.~\ref{fig:fit}, all data sets fitted to are compared with our
default NLO NRQCD results (solid lines).
For comparison, also the default results at LO (dashed lines) as well as
those of the CSM at NLO (dot-dashed lines) and LO (dotted lines) are shown.
In order to visualize the size of the NLO corrections to the hard-scattering
cross sections, the LO predictions are evaluated with the same LDMEs and PDFs.
The yellow and blue (shaded) bands indicate the theoretical errors on the
NLO NRQCD and CSM results.
We observe from Fig.~\ref{fig:fit} that the experimental data are nicely
described by NLO NRQCD, being almost exclusively contained within its error
bands, while they overshoot the NLO CSM predictions typically by 1--2 orders of
magnitude for hadroproduction and a factor of 3--5 for photoproduction.
In contrast to the LO analysis of Ref.~\refcite{Klasen:2001cu}, the
DELPHI data\cite{Abdallah:2003du} tend to systematically overshoot the NLO
NRQCD result, albeit the deviation is by no means significant in view of the
sizable experimental errors.
This may be attributed to the destructive interference of the $^1\!S_0^{[8]}$
and $^3\!P_J^{[8]}$ contributions, which is a genuine NLO phenomenon.
We have to bear in mind, however, that the DELPHI measurement comprises only 16
events with $p_T>1$~GeV and has not been confirmed by any of the other three
LEP~II experiments.
The Belle measurement,
$\sigma(e^+e^-\to J/\psi+X)=(0.43\pm0.13)$~pb,\cite{Pakhlov:2009nj} is
compatible both with the NLO NRQCD and CSM results,
$(0.70\genfrac{}{}{0pt}{}{+0.35}{-0.17})$~pb and
$(0.24\genfrac{}{}{0pt}{}{+0.20}{-0.09})$~pb, respectively.
However, the measured cross section was actually obtained from a data sample
with the multiplicity of charged tracks in the events being larger than four,
and corrections for the effect of this requirement were not performed, so that
the value quoted in Ref.~\refcite{Pakhlov:2009nj} just gives a lower bound on
the cross section.

\section{\boldmath Further tests of NRQCD factorization in unpolarized $J/\psi$
production\unboldmath}

\begin{figure}[ph]
\begin{center}
\begin{tabular}{ccc}
\includegraphics[width=0.31\textwidth]{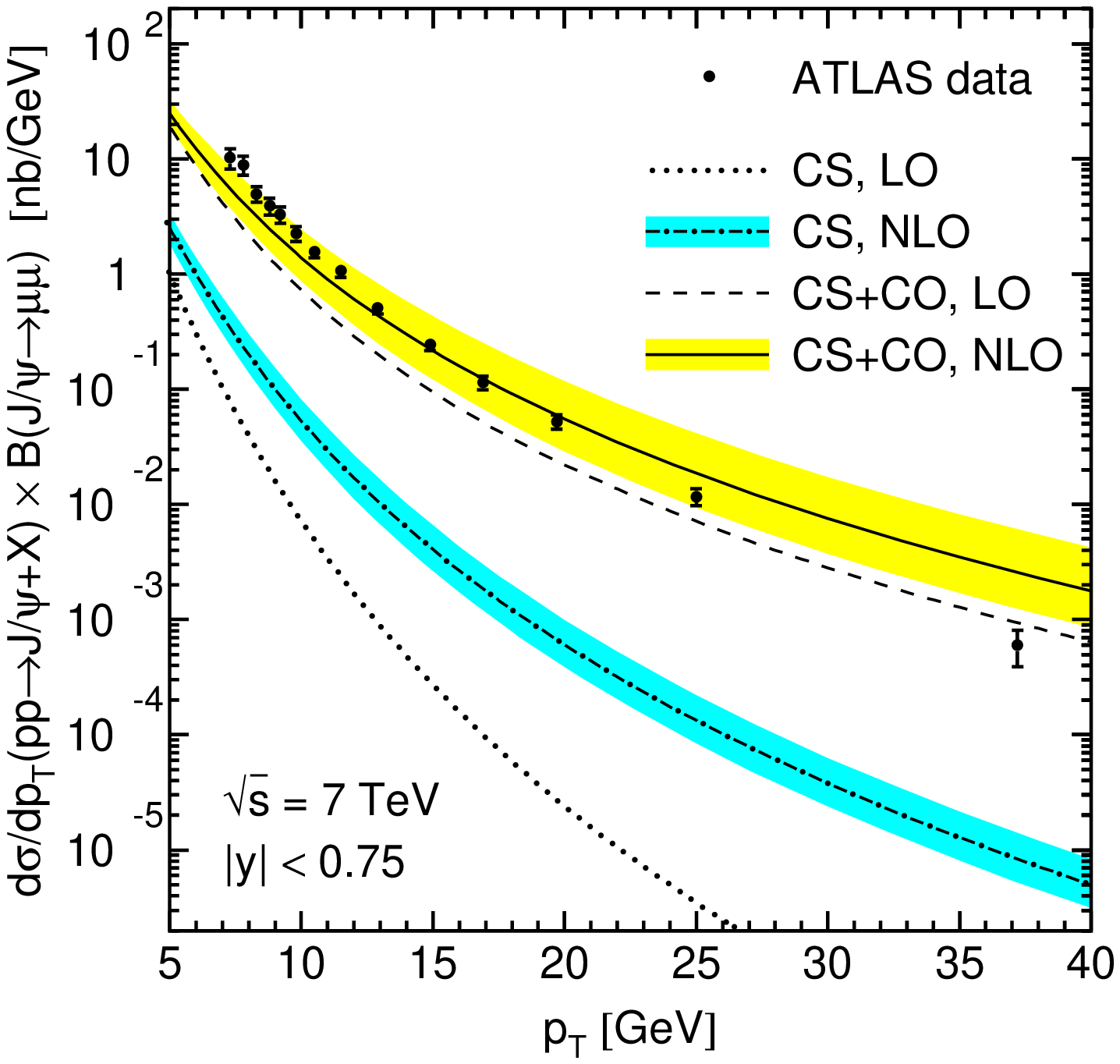}
&
\includegraphics[width=0.31\textwidth]{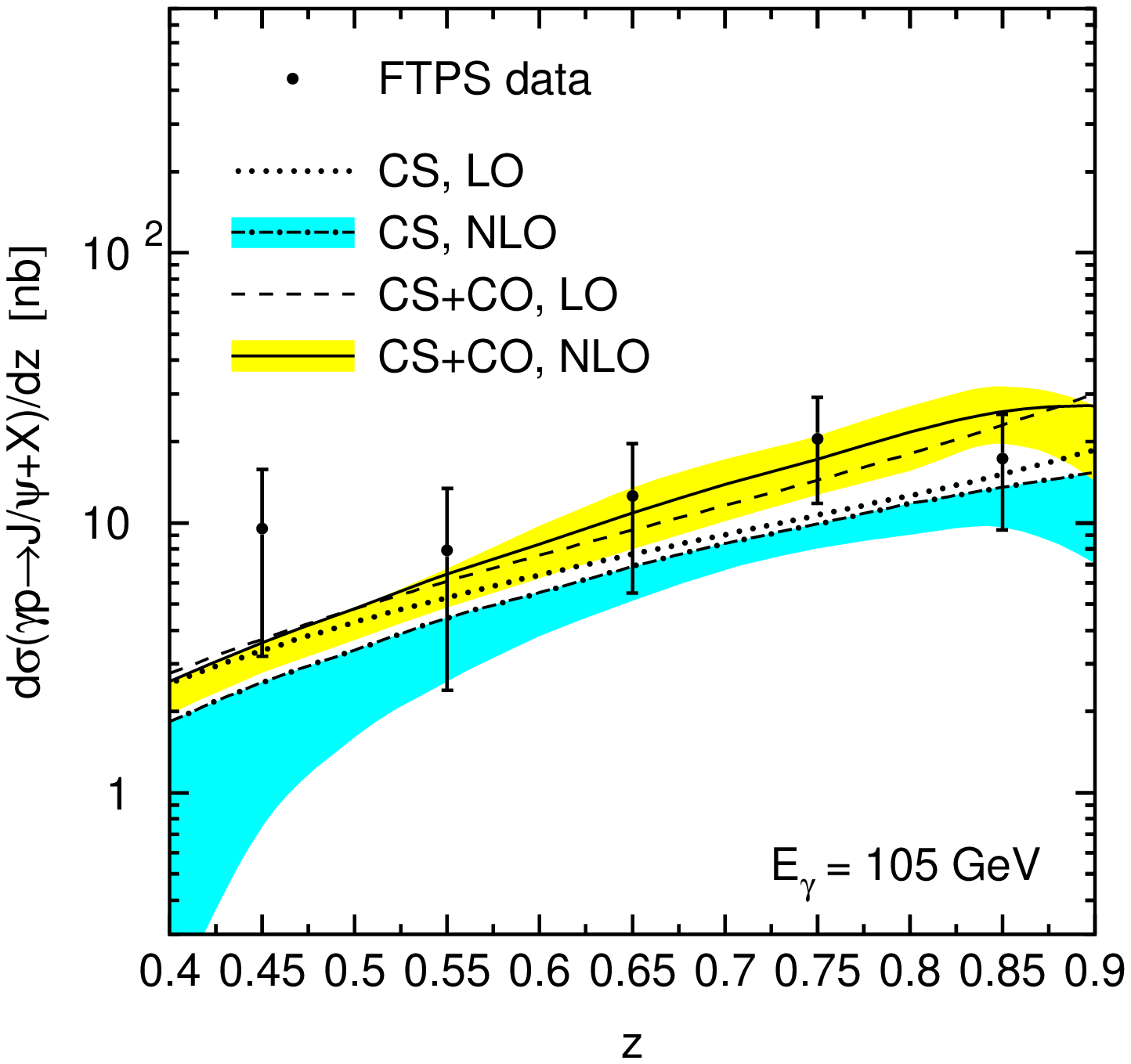}
&
\includegraphics[width=0.31\textwidth]{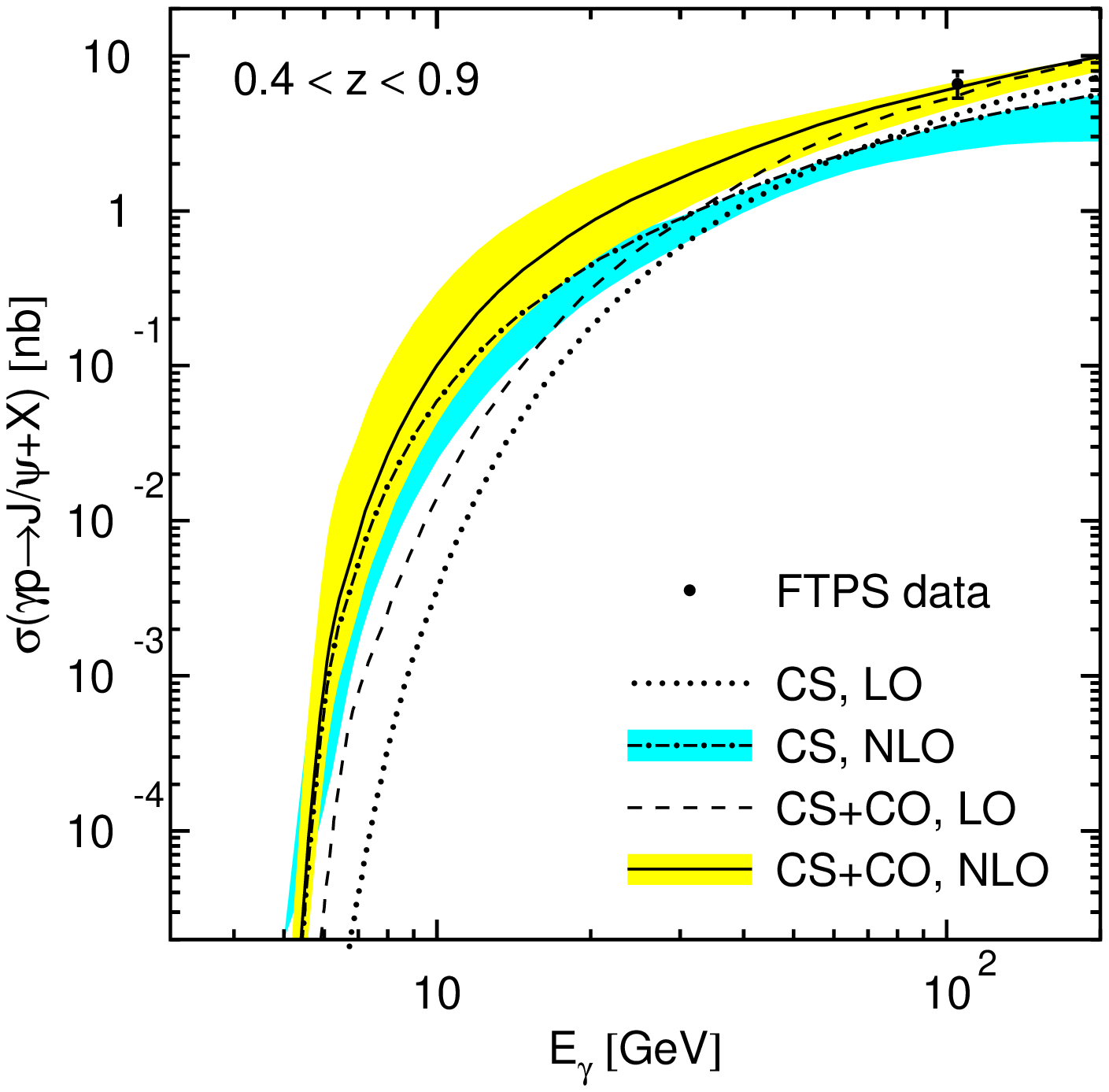}
\end{tabular}
\end{center}
\vspace*{8pt}
\caption{ATLAS data on $J/\psi$ inclusive
hadroproduction\protect\cite{Aad:2011sp} and FTPS data on $J/\psi$ inclusive
photoproduction in the fixed-target mode\protect\cite{Denby:1983az} compared
to NLO NRQCD predictions evaluated using set A of CO LDMEs from
Table~\protect\ref{tab:fit}.
\protect\label{fig:atlasftps}}
\end{figure}

\begin{figure}[ph]
\begin{center}
\includegraphics[width=0.75\textwidth]{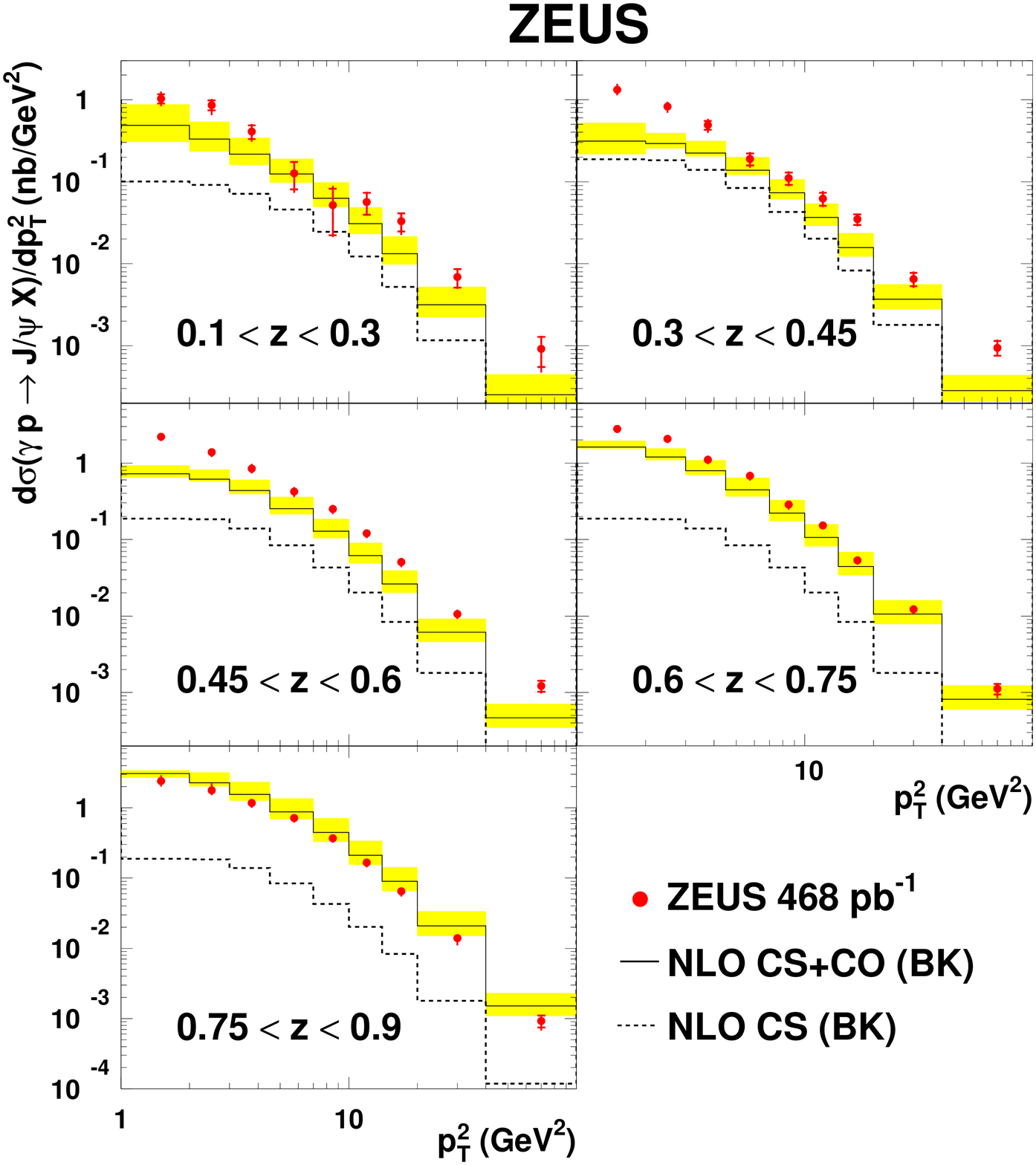}
\end{center}
\vspace*{8pt}
\caption{ZEUS data on $J/\psi$ inclusive
photoproduction\protect\cite{Abramowicz:2012dh} compared to NLO NRQCD
predictions evaluated using set A of CO LDMEs from Table~\protect\ref{tab:fit}.
\protect\label{fig:zeus}}
\end{figure}

Three data sets not included in the global fit,\cite{Butenschoen:2011yh} from
hadroproduction and from photoproduction in the fixed-target and colliding-beam
modes, are nicely reproduced by our NLO NRQCD predictions, as may be seen from
Figs.~\ref{fig:atlasftps} and \ref{fig:zeus}.
They were taken by the ATLAS Collaboration\cite{Aad:2011sp} at the LHC, by
Denby et al.\cite{Denby:1983az} at the Fermilab Tagged-Photon Spectrometer,
and by the ZEUS Collaboration\cite{Abramowicz:2012dh} at HERA~II.
The $\chi_{\rm d.o.f.}^2$ values evaluated using our default NLO NRQCD
predictions read 10.74,\footnote{%
This value is reduced to 4.88 if the data point at the largest value of $p_T$
is omitted.}
0.40, and 7.50, respectively.
We conclude that NRQCD factorization passes this nontrivial test, which, in the
case of Refs.~\refcite{Aad:2011sp,Denby:1983az}, probes kinematic regions far
outside those covered by the global fit.\cite{Butenschoen:2011yh} 

\section{\boldmath $J/\psi$ polarization\unboldmath}

The polarization of the $J/\psi$ meson is conveniently analyzed experimentally
by measuring the angular distribution of its leptonic decays, which is
customarily parametrized using the three polarization observables
$\lambda_\theta$, $\lambda_\phi$, and $\lambda_{\theta\phi}$,
as\cite{Lam:1978pu}
\begin{equation}
W(\theta,\phi)\propto 1+\lambda_\theta\cos^2\theta
+\lambda_\phi\sin^2\theta\cos(2\phi)
+\lambda_{\theta\phi}\sin(2\theta)\cos\phi,
\end{equation}
where $\theta$ and $\phi$ are respectively the polar the azimuthal angles of
$l^+$ in the $J/\psi$ rest frame.
This definition depends on the choice of coordinate frame.
In the experimental
analyses,\cite{Aaron:2010gz,Chekanov:2009ad,Affolder:2000nn,Abulencia:2007us,Abelev:2011md}
the helicity (recoil), Collins-Soper, and target frames were employed, in which
the polar axes point in the directions of $-(\vec{p}_p+\vec{p}_{\overline{p}})$,
$\vec{p}_p/|\vec{p}_p|-\vec{p}_{\overline{p}}/|\vec{p}_{\overline{p}}|$, and
$-\vec{p}_p$, respectively.
The values $\lambda_\theta=0,+1,-1$ correspond to unpolarized, fully
transversely polarized, and fully longitudinally polarized $J/\psi$ mesons,
respectively.
The alternative notation $\lambda=\lambda_\theta$, $\mu=\lambda_{\theta\phi}$,
and $\nu=2\lambda_\phi$ is frequently encountered in the literature.
In Refs.~\refcite{Affolder:2000nn,Abulencia:2007us}, $\lambda_\theta$ is called
$\alpha$.

Working in the spin density matrix formalism and denoting the $z$ component of
$S$ by $i,j=0,\pm1$, we have
\begin{equation}
\lambda_\theta=\frac{d\sigma_{11}-d\sigma_{00}}{d\sigma_{11}+d\sigma_{00}},
\qquad
\lambda_\phi=\frac{d\sigma_{1,-1}}{d\sigma_{11}+d\sigma_{00}},
\qquad
\lambda_{\theta\phi}=
\frac{\sqrt{2}\re d\sigma_{10}}{d\sigma_{11}+d\sigma_{00}},
\label{eq:lam}
\end{equation}
where $d\sigma_{ij}$ is the $ij$ component of the differential cross section.
An expression of $d\sigma_{ij}$ in terms of PDFs and partonic spin density
matrix elements may be found in Eq.~(3) of Ref.~\refcite{Butenschoen:2011ks}.

\begin{figure}[ph]
\begin{center}
\begin{tabular}{ccc}
\includegraphics[width=0.31\textwidth]{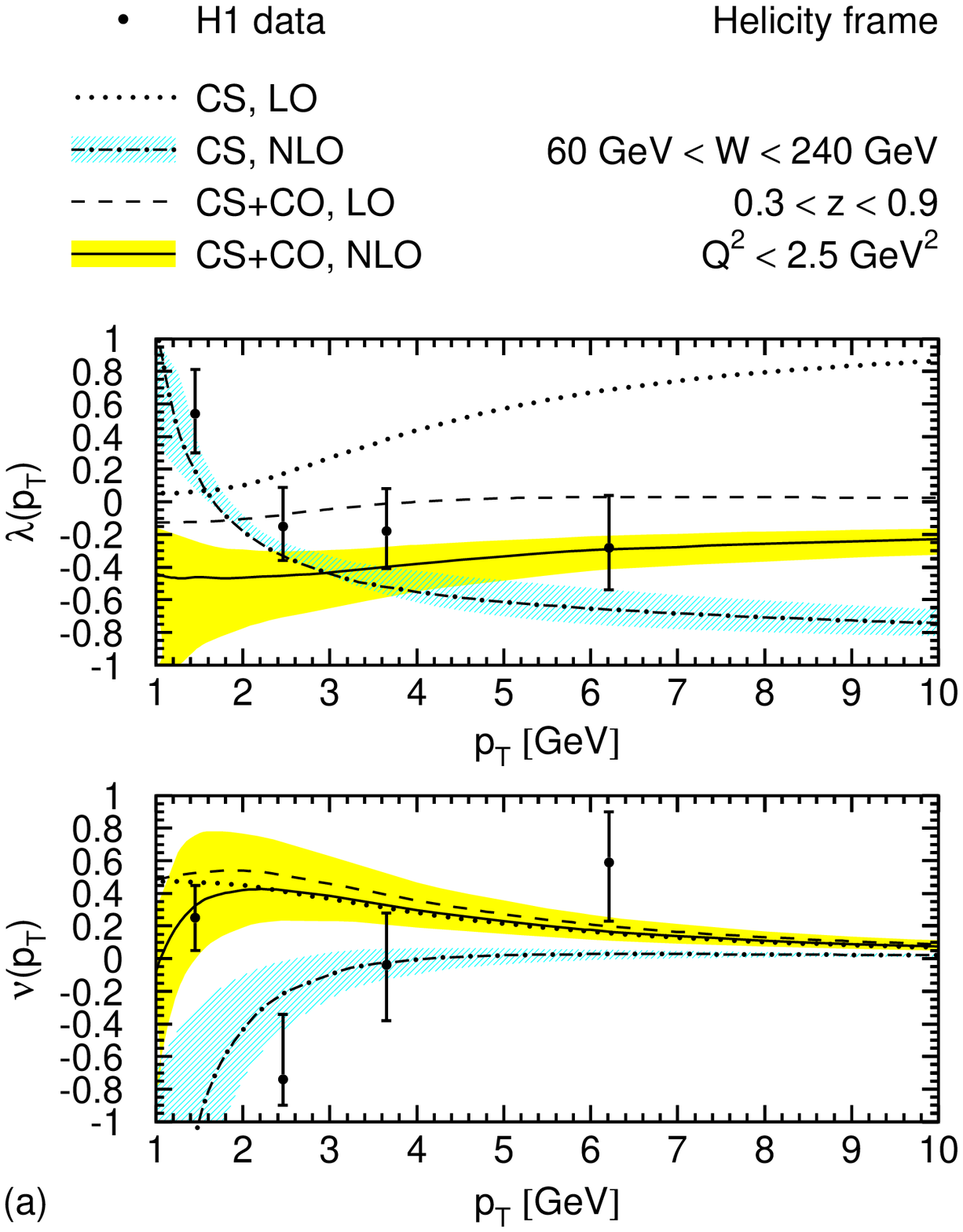}
&
\includegraphics[width=0.31\textwidth]{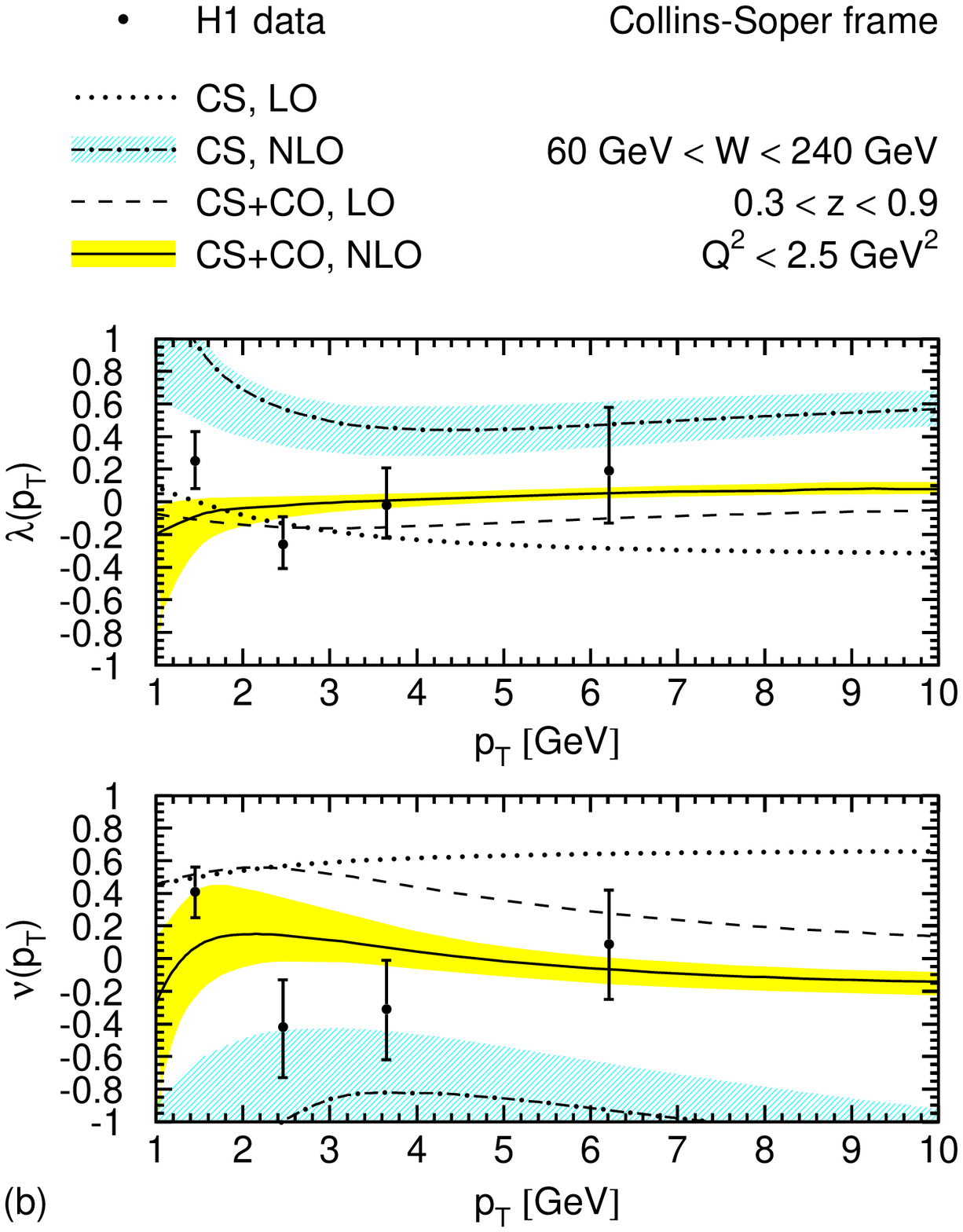}
&
\includegraphics[width=0.31\textwidth]{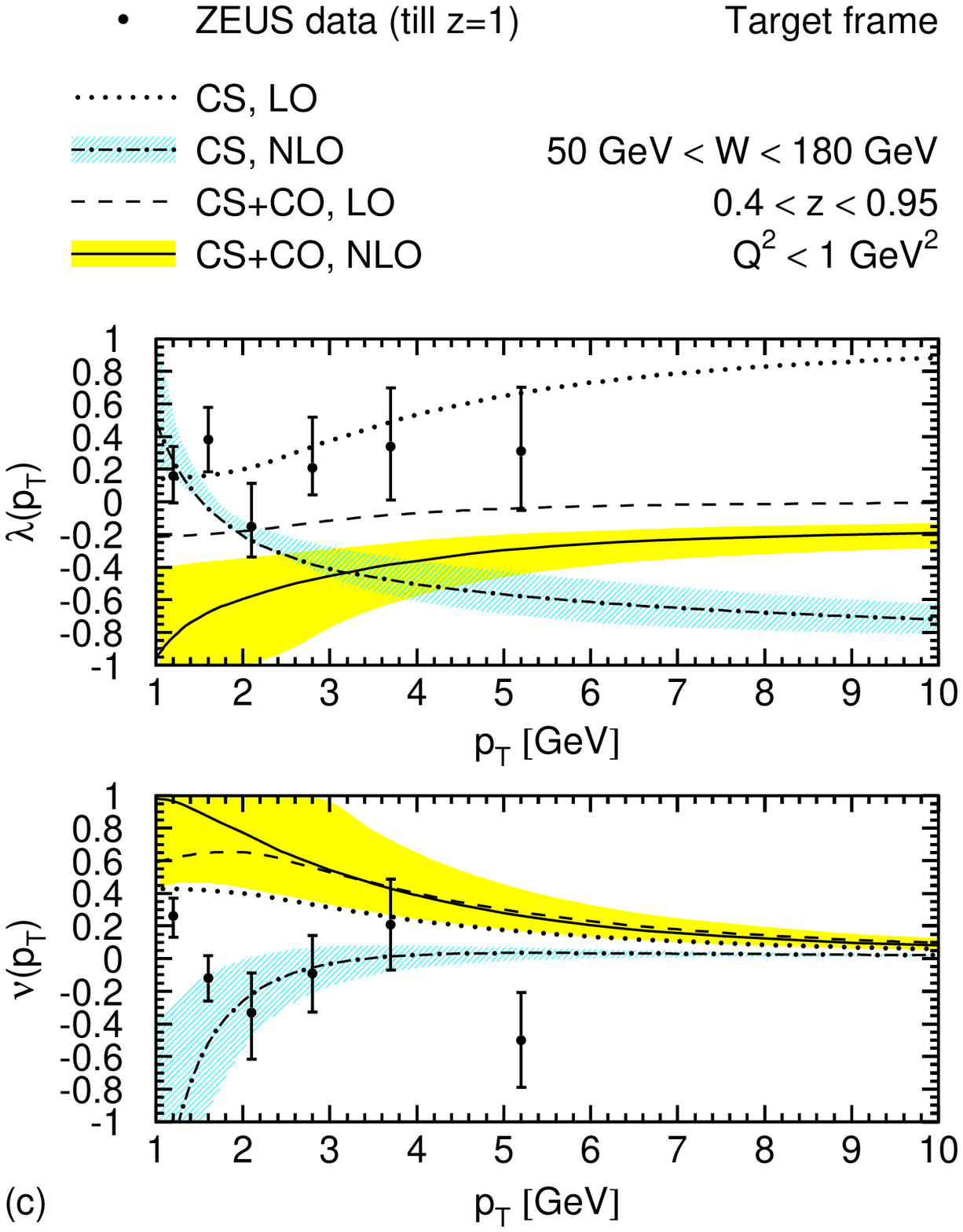}
\\
\includegraphics[width=0.31\textwidth]{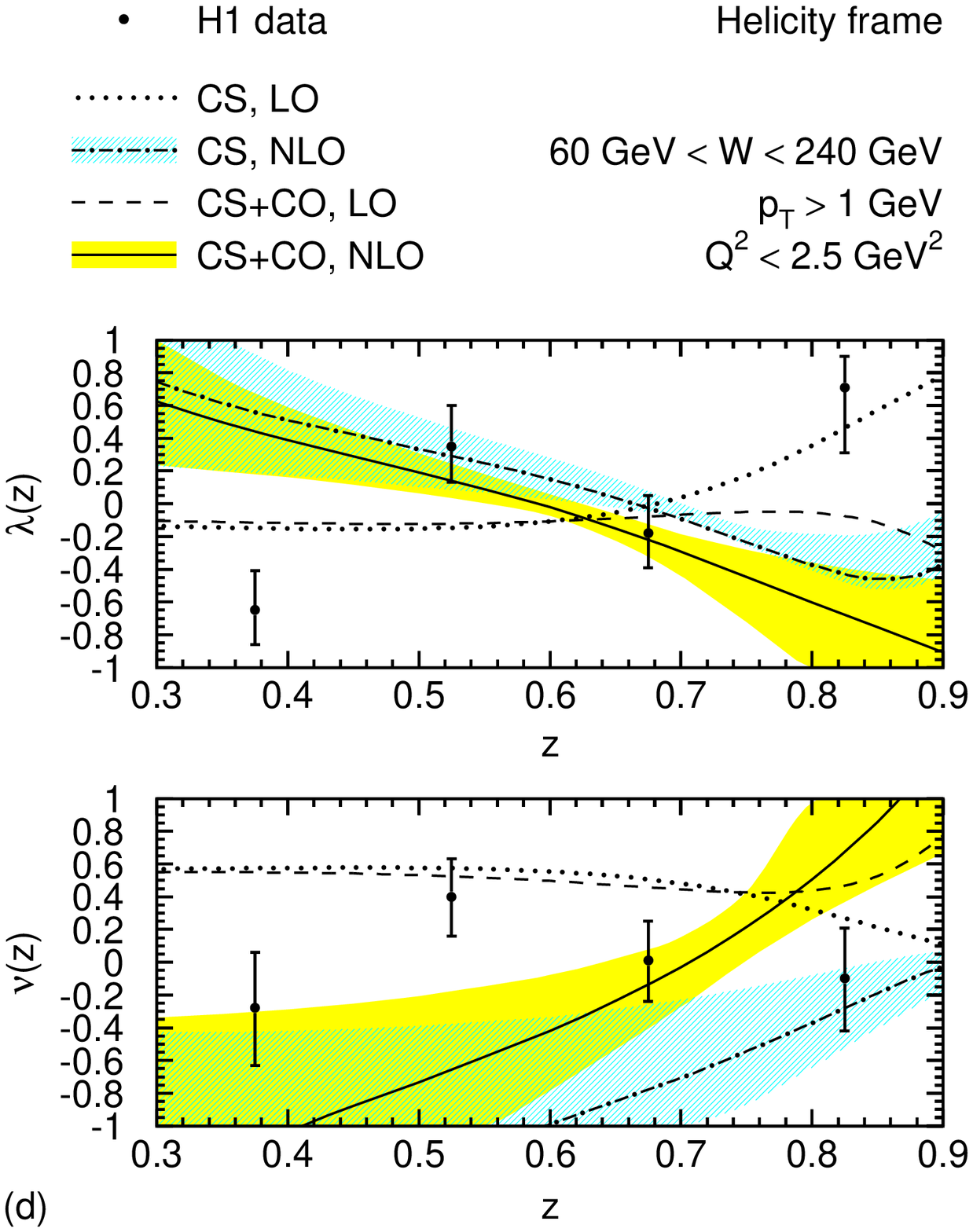}
&
\includegraphics[width=0.31\textwidth]{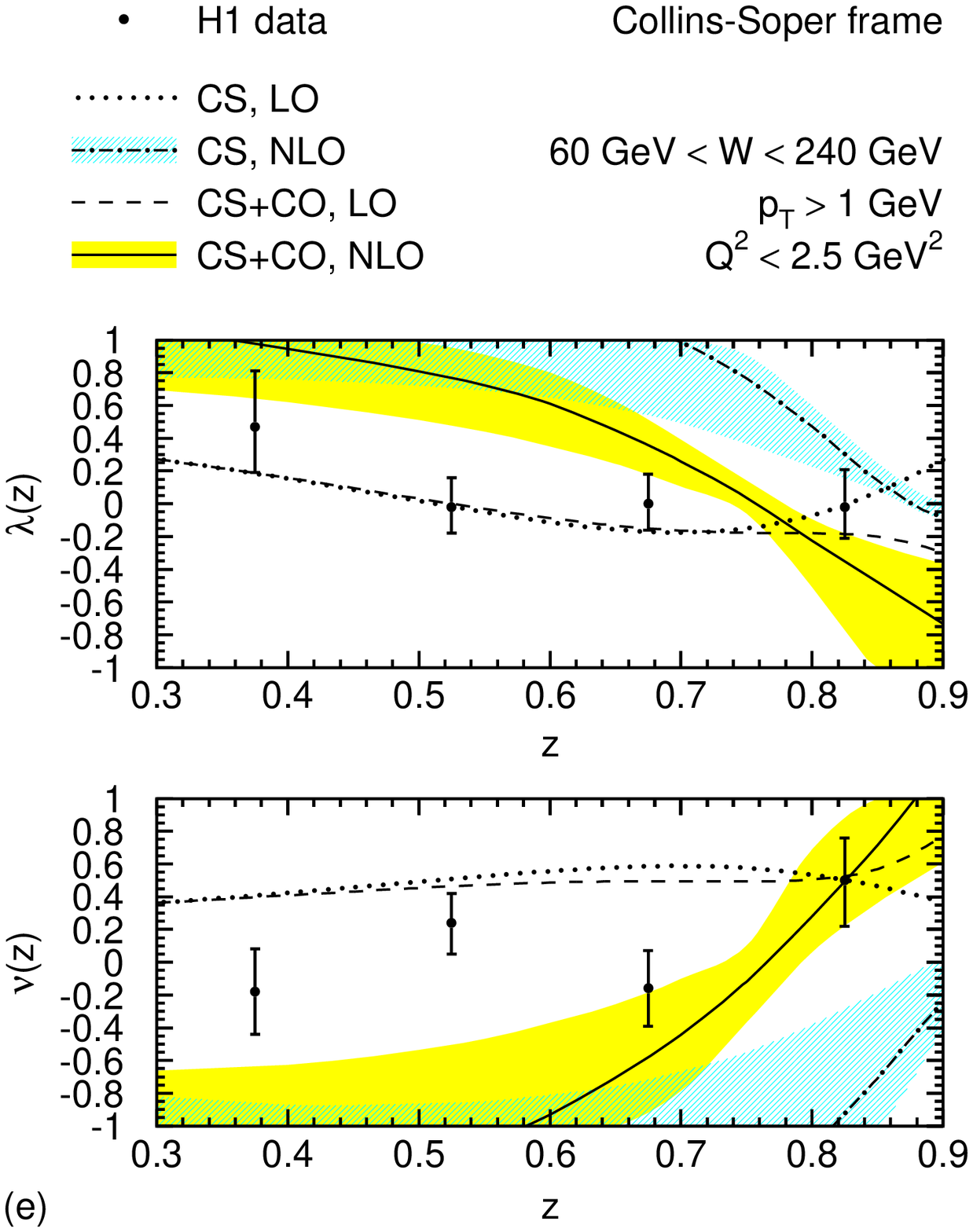}
&
\includegraphics[width=0.31\textwidth]{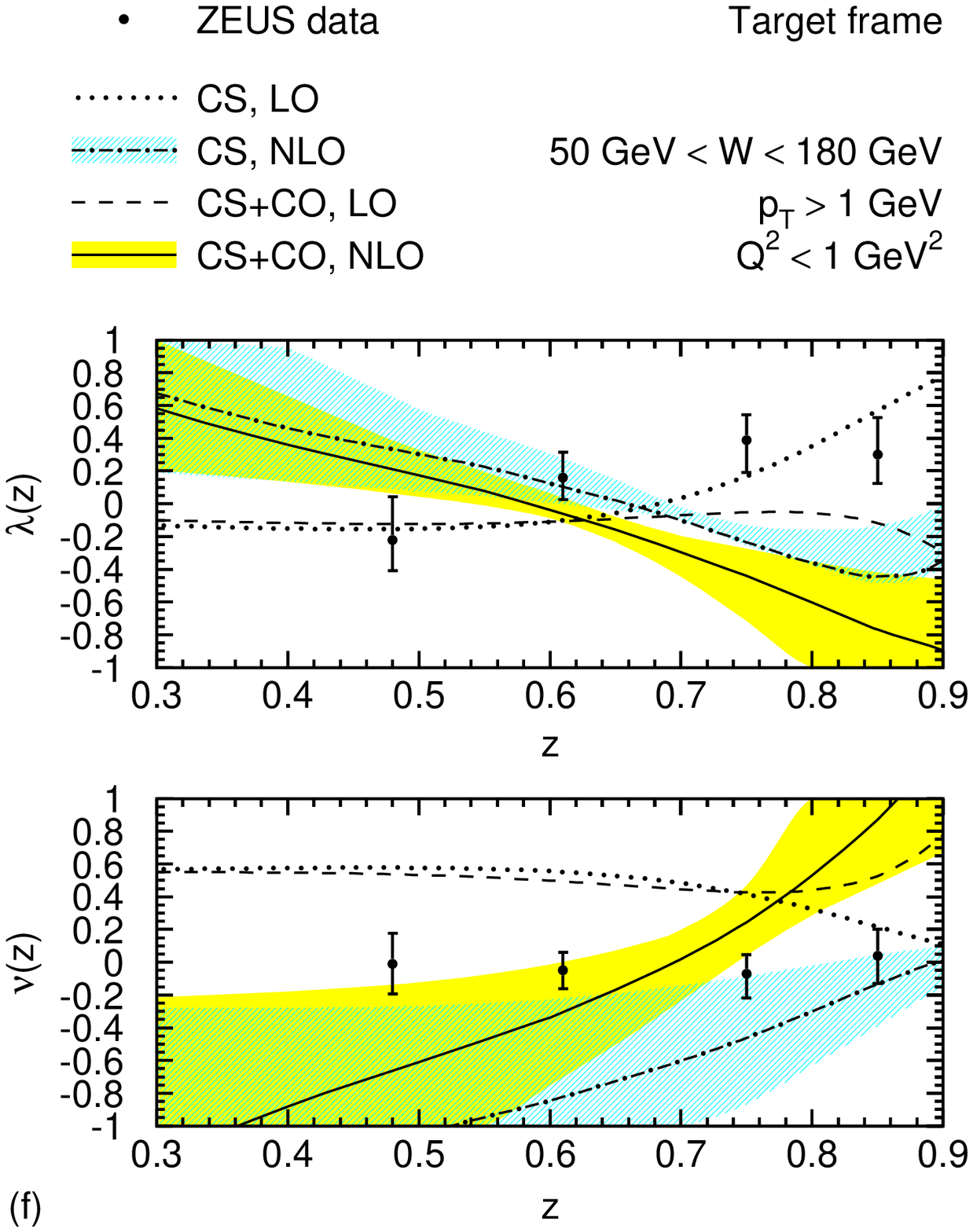}
\end{tabular}
\end{center}
\vspace*{8pt}
\caption{The polarization parameters $\lambda$ and $\nu$ for direct
photoproduction at HERA evaluated at NLO in the CSM and in
NRQCD\protect\cite{Butenschoen:2011ks} using set B of CO LDME from
Table~\protect\ref{tab:fit} are compared to H1\protect\cite{Aaron:2010gz} and
ZEUS\protect\cite{Chekanov:2009ad} data.
The theoretical uncertainties are due to scale variations in the CSM
(blue bands) and include also the errors on the CO LDMEs (yellow bands) in
NRQCD.
\protect\label{fig:PhotoPol}}
\end{figure}

Our results for direct photoproduction\cite{Butenschoen:2011ks} are shown in
Fig.~\ref{fig:PhotoPol}.
We compare our NLO predictions for the parameters $\lambda$ and $\nu$ as
functions of $p_T$ and $z$ with measurements by the H1
Collaboration\cite{Aaron:2010gz} in the helicity and Collins-Soper frames and
by the ZEUS Collaboration\cite{Chekanov:2009ad} in the target frame.
Unfortunately, the H1\cite{Aaron:2010gz} and ZEUS\cite{Chekanov:2009ad} data
do not yet allow us to distinguish the production mechanisms clearly.
However, kinematical regions can be identified in which a clear distinction
could be possible in more precise experiments at a future $ep$ collider, such
as the CERN LHeC.\cite{Armesto:2012jn}
At higher values of $p_T$, NRQCD predicts the $J/\psi$ meson to be largely
unpolarized, in contrast to the CSM.
In the $z$ distributions, the scale uncertainties are sizable, and the error
bands of the CSM and NRQCD predictions largely overlap.

\begin{figure}[ph]
\begin{center}
\begin{tabular}{ccc}
\includegraphics[width=0.31\textwidth]{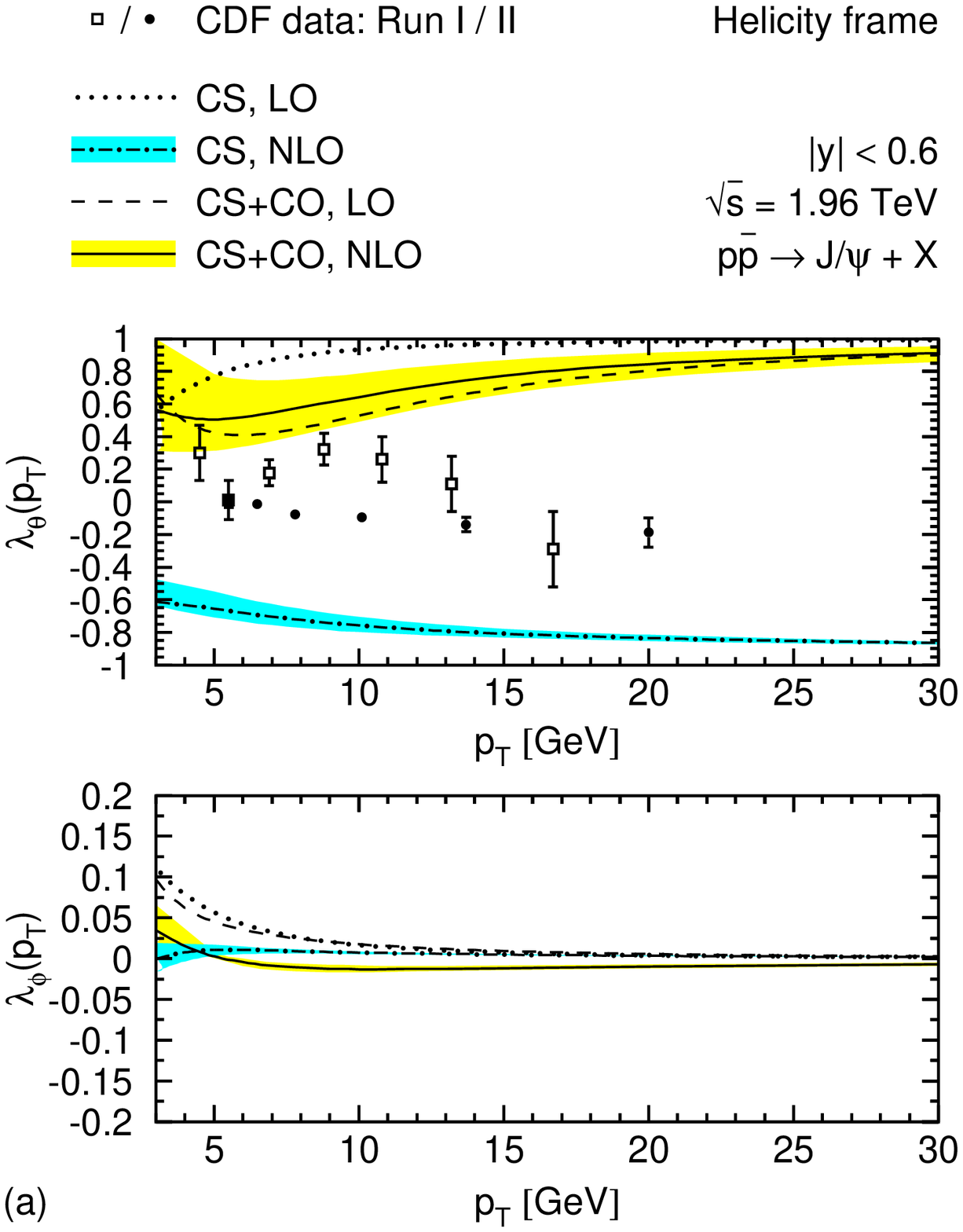}
&
\includegraphics[width=0.31\textwidth]{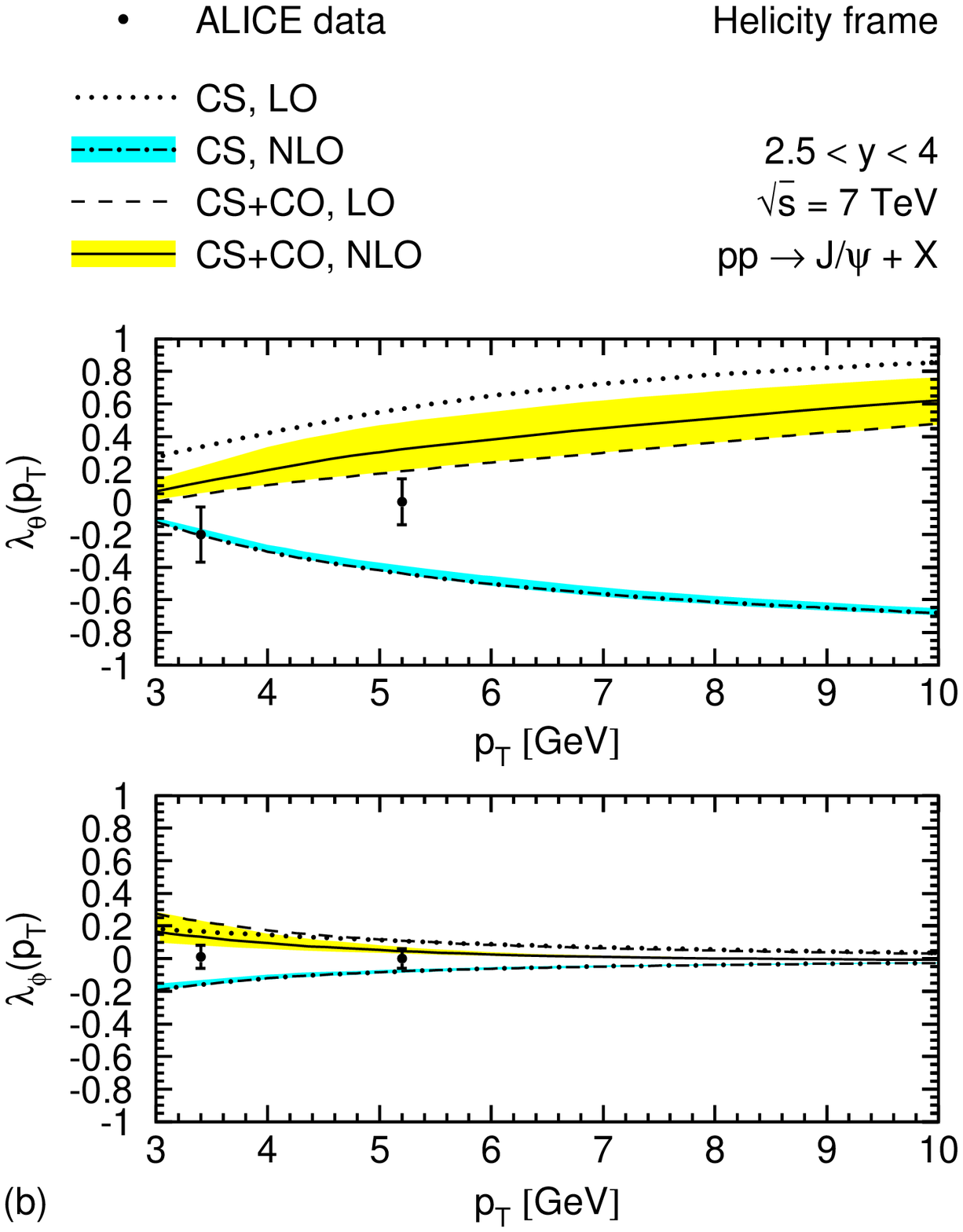}
&
\includegraphics[width=0.31\textwidth]{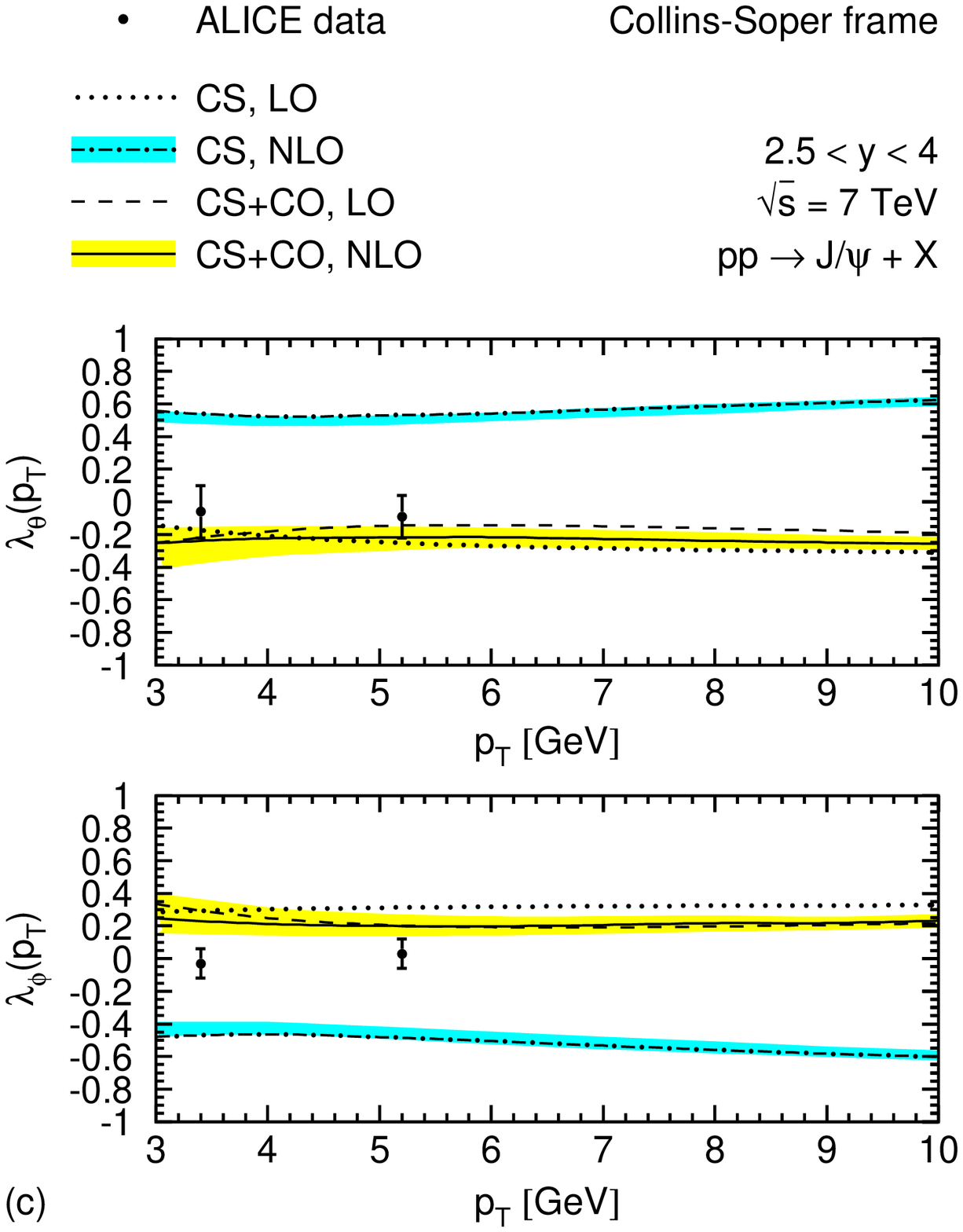}
\end{tabular}
\end{center}
\vspace*{8pt}
\caption{The polarization parameters $\lambda_\theta$ and $\lambda_\phi$ for
hadroproduction evaluated at NLO in the CSM and in
NRQCD\protect\cite{Butenschoen:2012px,Butenschoen:2012qh} using set B of CO
LDME from Table~\protect\ref{tab:fit} are compared to CDF data from Tevatron
runs I\protect\cite{Affolder:2000nn} and II\protect\cite{Abulencia:2007us} and
to ALICE data.\protect\cite{Abelev:2011md}
The theoretical uncertainties are due to scale variations in the CSM
(blue bands) and include also the errors on the CO LDMEs (yellow bands) in
NRQCD.
\protect\label{fig:HadroPol}}
\end{figure}

Our results for direct
hadroproduction\cite{Butenschoen:2012px,Butenschoen:2012qh} are shown in
Fig.~\ref{fig:HadroPol}.
We compare our predictions for the parameters $\lambda_\theta$ and
$\lambda_\phi$ as functions of $p_T$ in the helicity and Collins-Soper frames
with the measurements by CDF\cite{Affolder:2000nn,Abulencia:2007us} and
ALICE.\cite{Abelev:2011md}
In the helicity frame, the CSM predicts the $J/\psi$ polarization to be
strongly longitudinal at NLO, while NRQCD predicts it to be strongly
transverse.
In the Collins-Soper frame, the situation is inverted.
The precise CDF measurement at Tevatron run~II,\cite{Abulencia:2007us} which
is partially in disagreement with the one at run~I,\cite{Affolder:2000nn}
found the $J/\psi$ mesons to be largely unpolarized in the helicity frame,
which is in contradiction with both the CSM and NRQCD predictions at NLO.
The early ALICE data\cite{Abelev:2011md} is, however, compatible with NRQCD at
NLO, favoring NRQCD over the CSM.

\section{Comparisons with the literature}

After our NLO NRQCD studies of $J/\psi$
polarization,\cite{Butenschoen:2011ks,Butenschoen:2012px,Butenschoen:2012qh}
two others appeared, which are, however, limited to hadroproduction.
In Ref.~\refcite{Chao:2012iv}, it was shown that the measured hadroproduction
cross sections and the CDF~II polarization measurement can be simultaneously
described by NRQCD at NLO with one of the three CO LDME sets listed in the
fourth column of Table~\ref{tab:com}.
In Ref.~\refcite{Gong:2012ug}, the polarization of promptly produced $J/\psi$
mesons was studied by also including the feed-down from polarized $\chi_{cJ}$
and $\psi^\prime$ mesons as described in
Refs.~\refcite{Braaten:1999qk,Kniehl:2000nn}.
To this end, the CO LDMEs of the $\chi_{cJ}$ and $\psi^\prime$ mesons were
fitted to LHCb (and CDF) unpolarized production data, and the resulting cascade
decay rates into $J/\psi$ mesons were then used as feed-down contributions to
determine the $J/\psi$ CO LDMEs in a fit to unpolarized $J/\psi$ production
data from LHCb and CDF with $p_T>7$~GeV.
The resulting LDMEs may be found in the third column of Table~\ref{tab:com}. 
Reference~\refcite{Gong:2012ug} predicts the $J/\psi$ polarization to be
moderately transverse in the helicity frame.

\begin{figure}[ph]
\begin{center}
\begin{tabular}{cccc}
\includegraphics[width=0.22\textwidth]{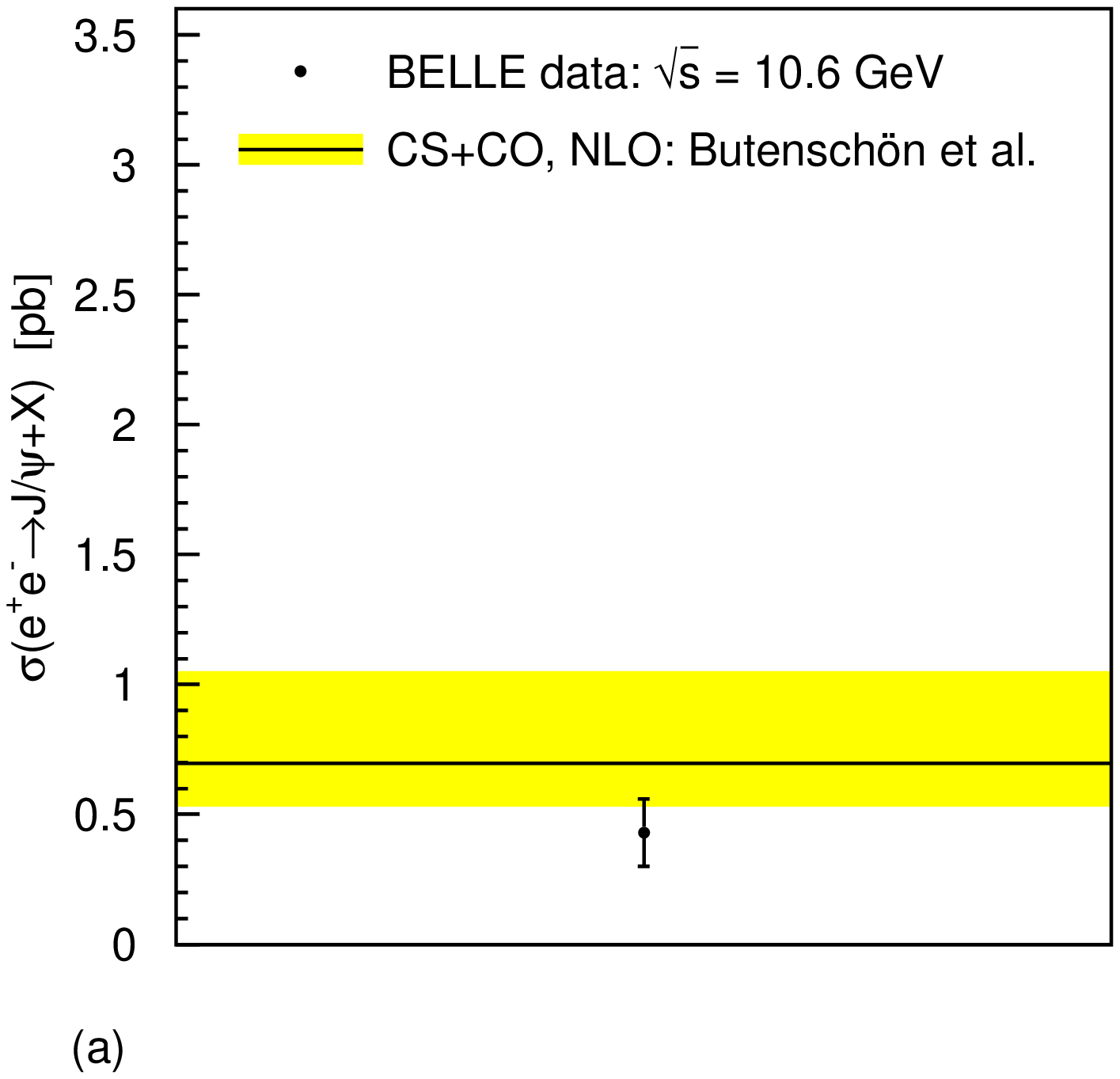}
&
\includegraphics[width=0.22\textwidth]{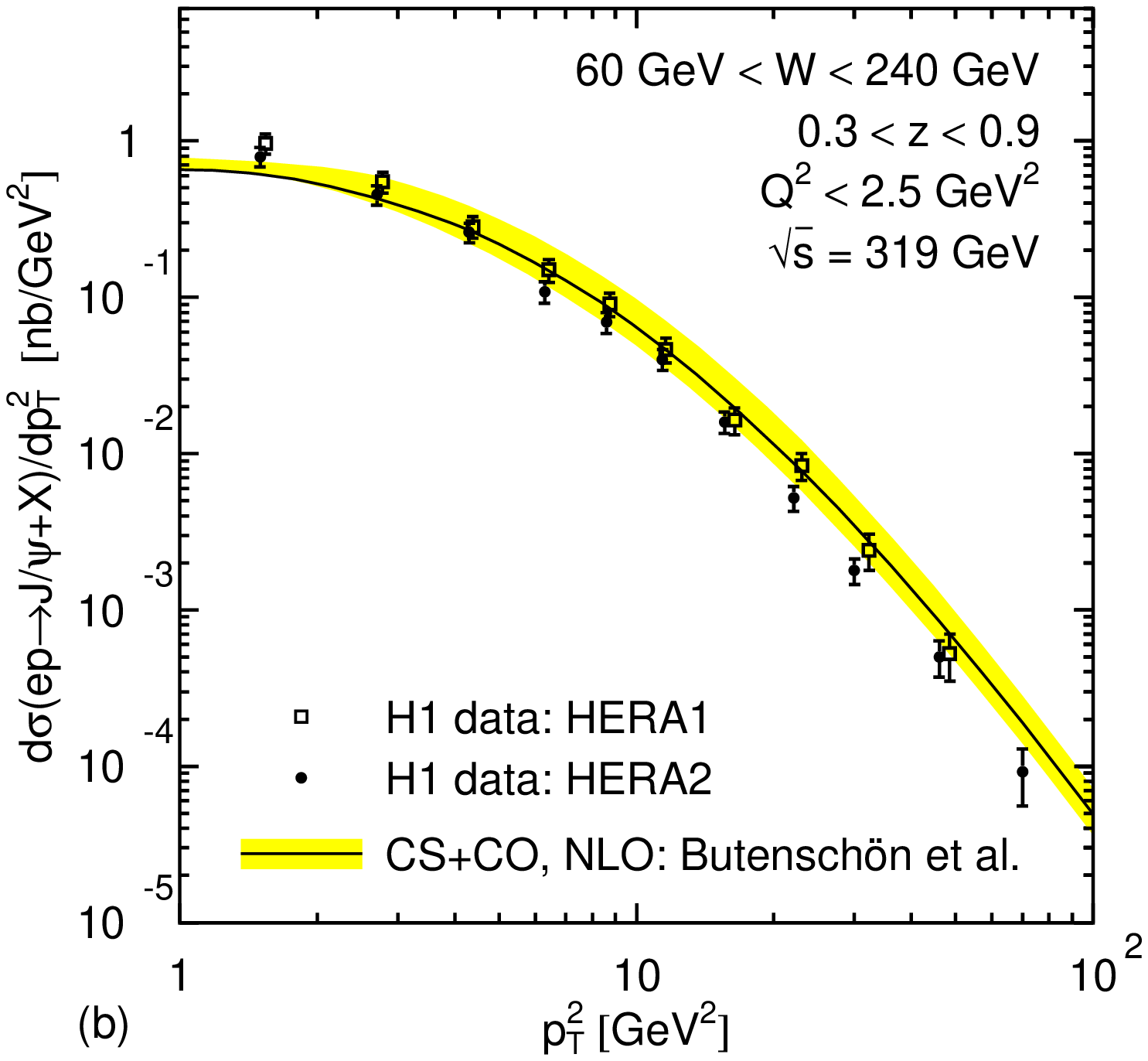}
&
\includegraphics[width=0.22\textwidth]{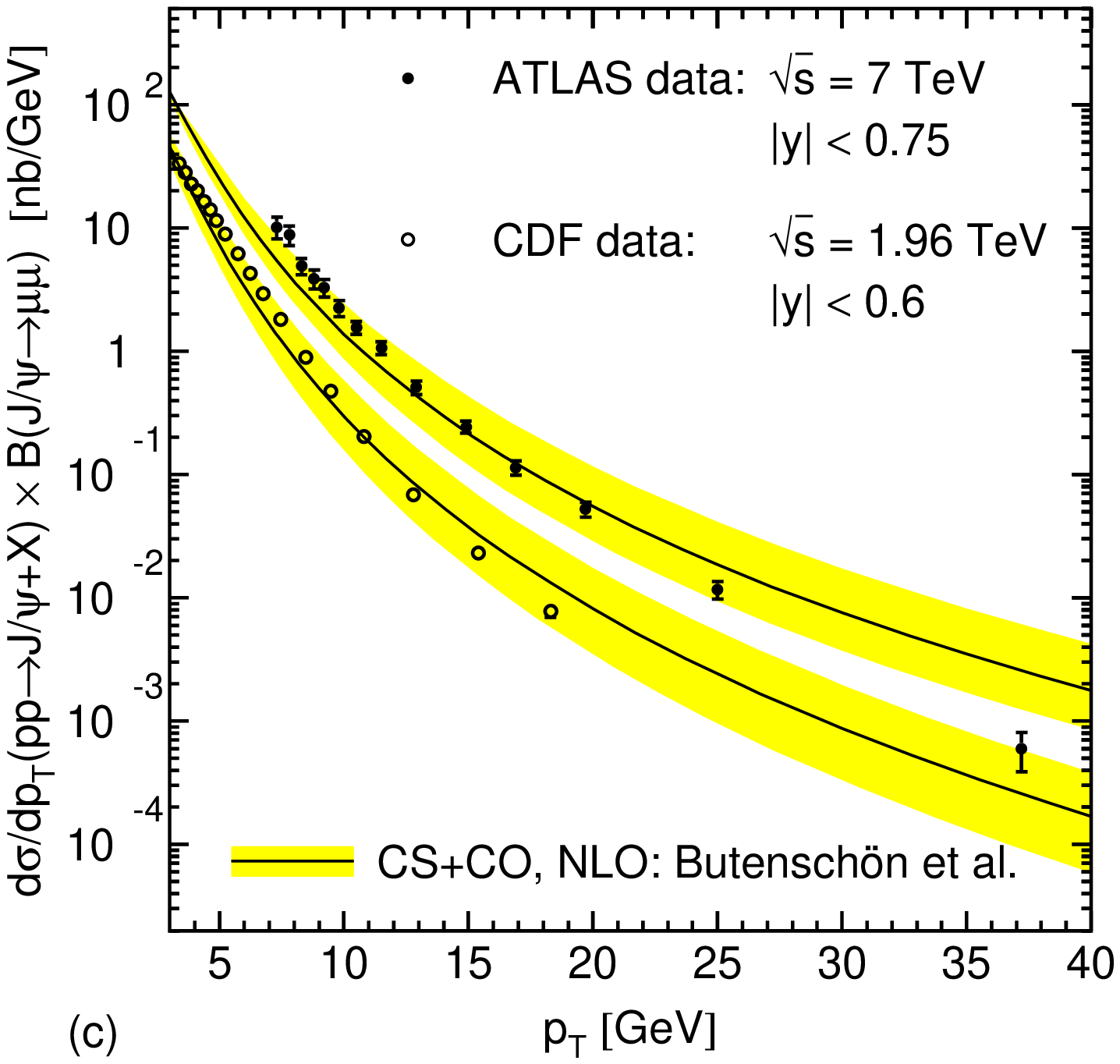}
&
\includegraphics[width=0.22\textwidth]{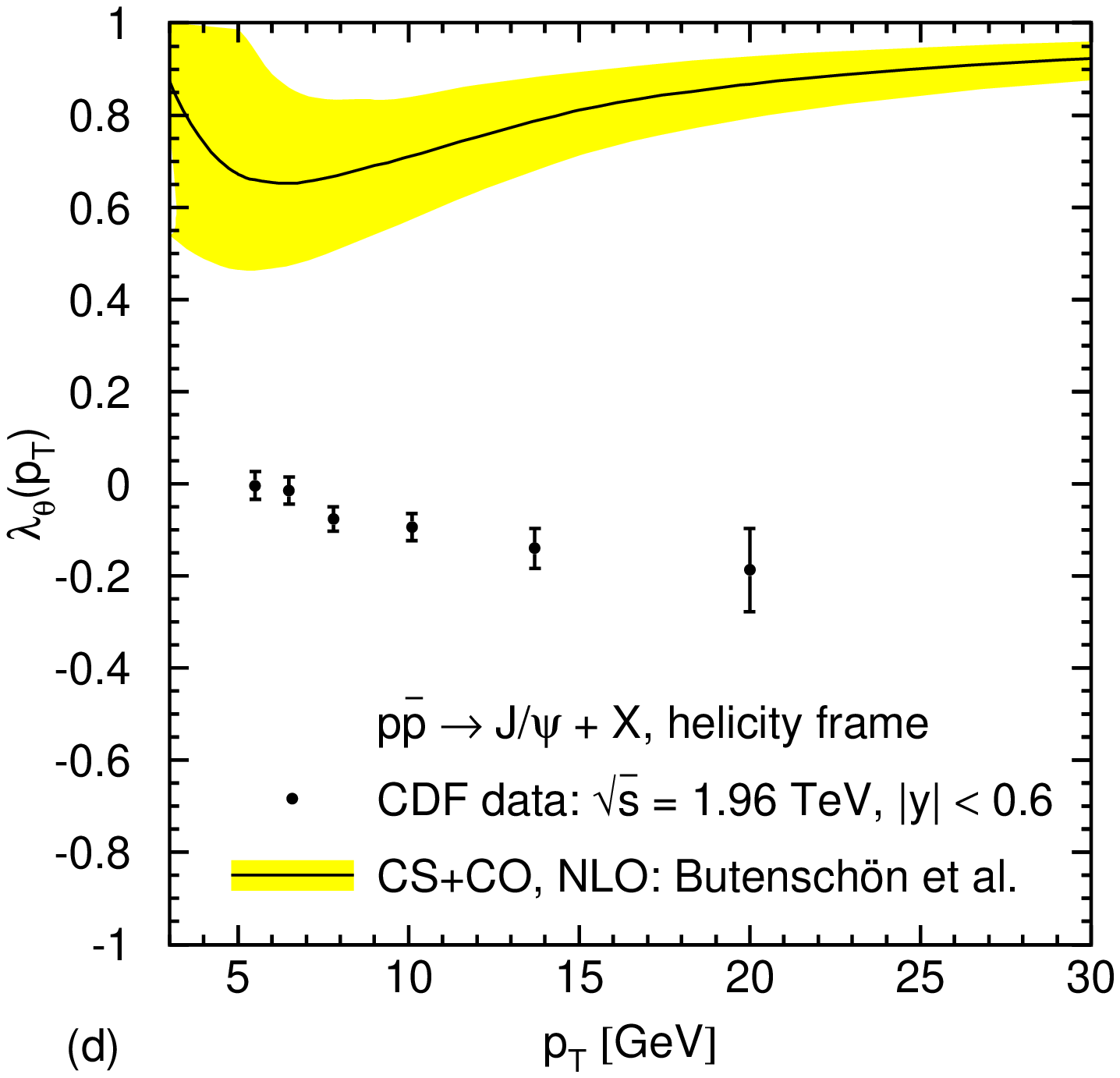}
\\
\includegraphics[width=0.22\textwidth]{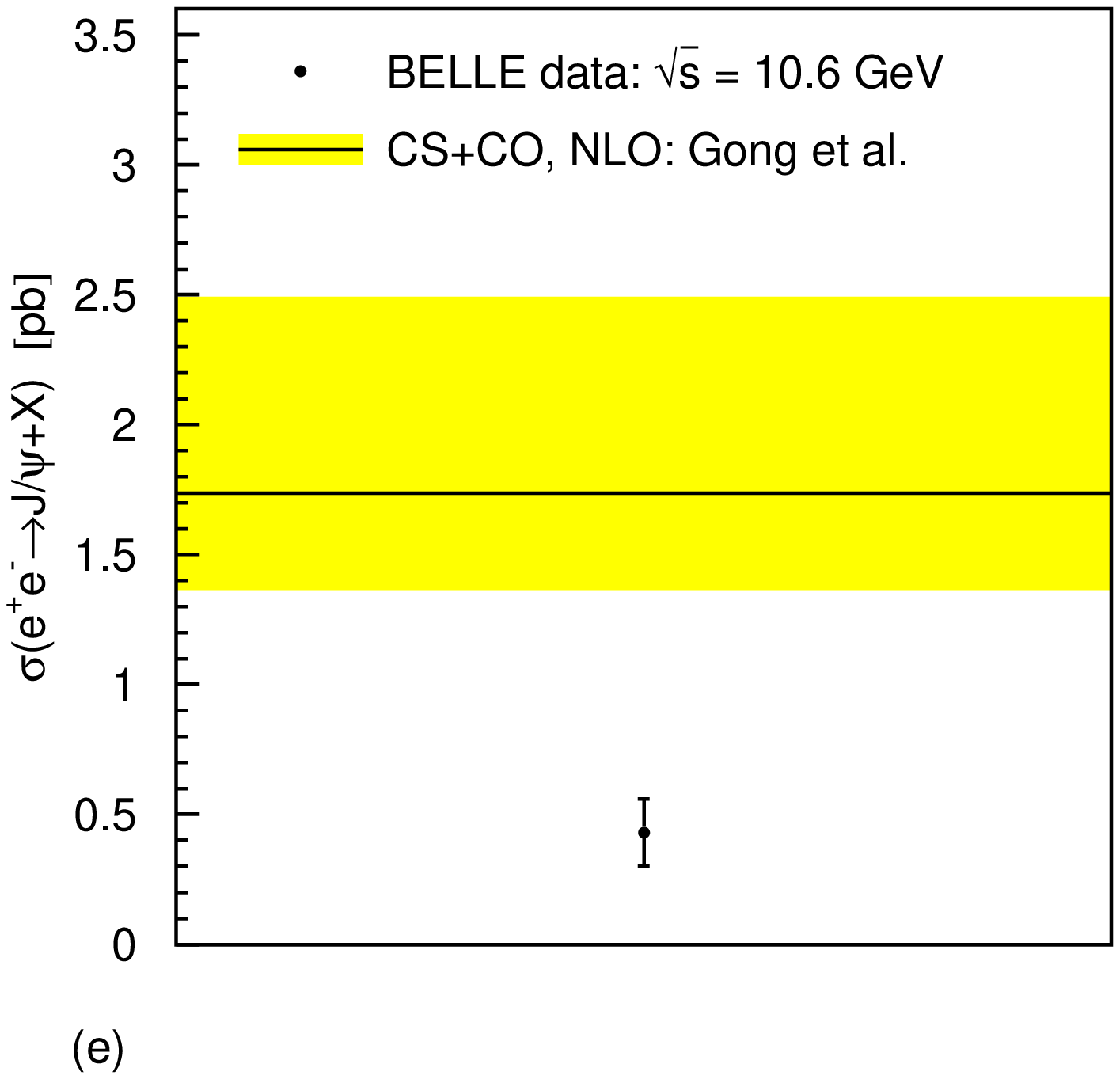}
&
\includegraphics[width=0.22\textwidth]{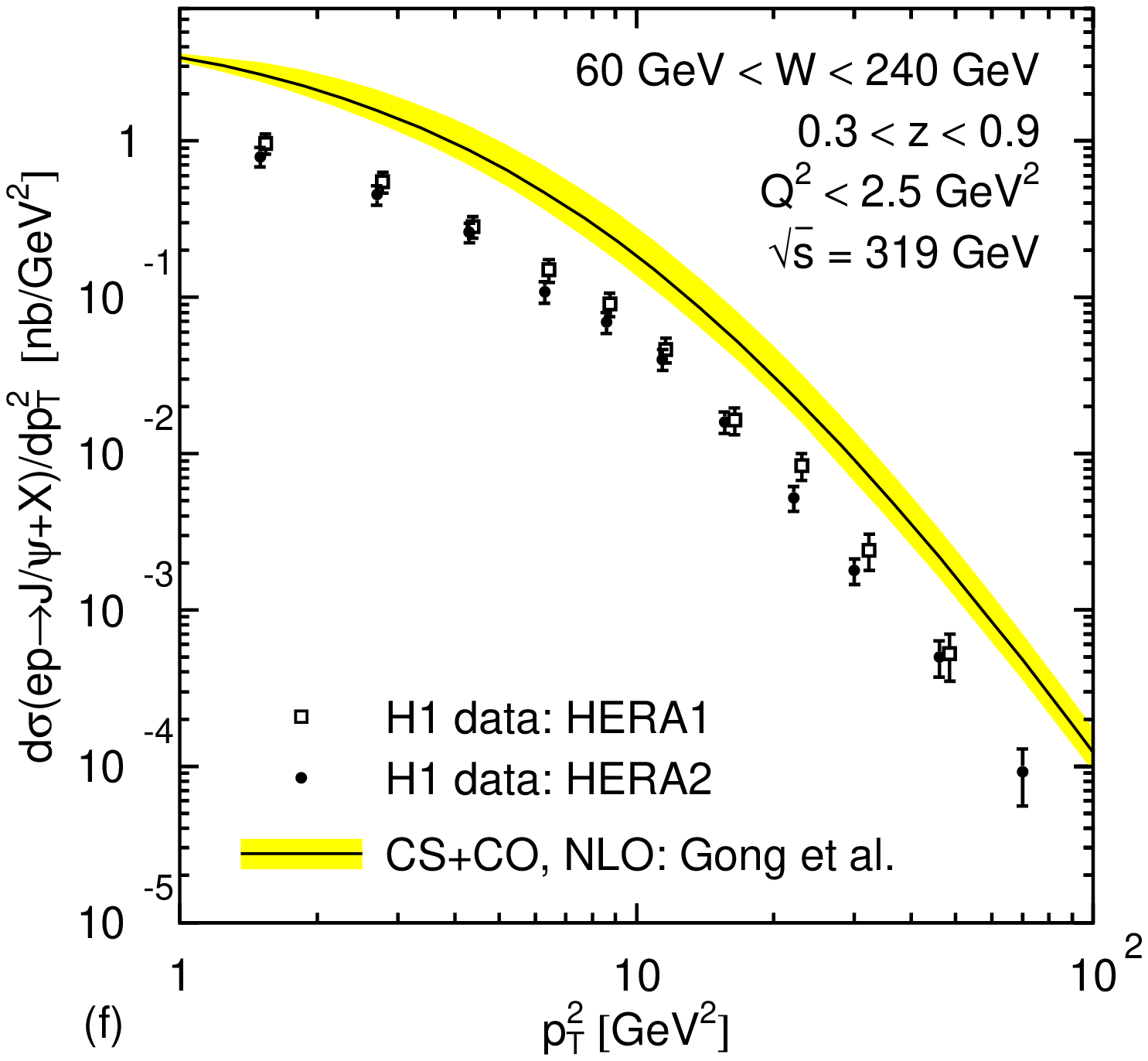}
&
\includegraphics[width=0.22\textwidth]{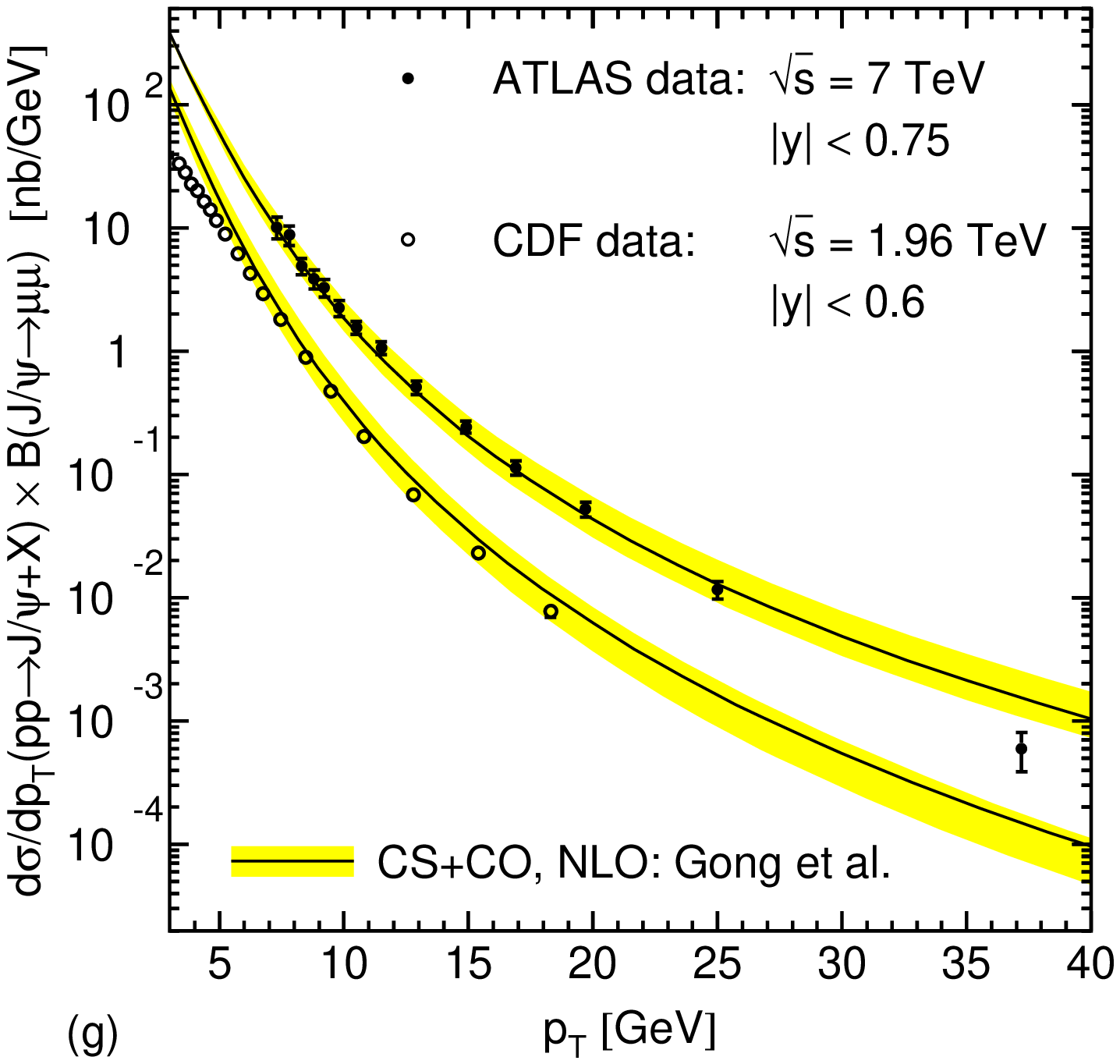}
&
\includegraphics[width=0.22\textwidth]{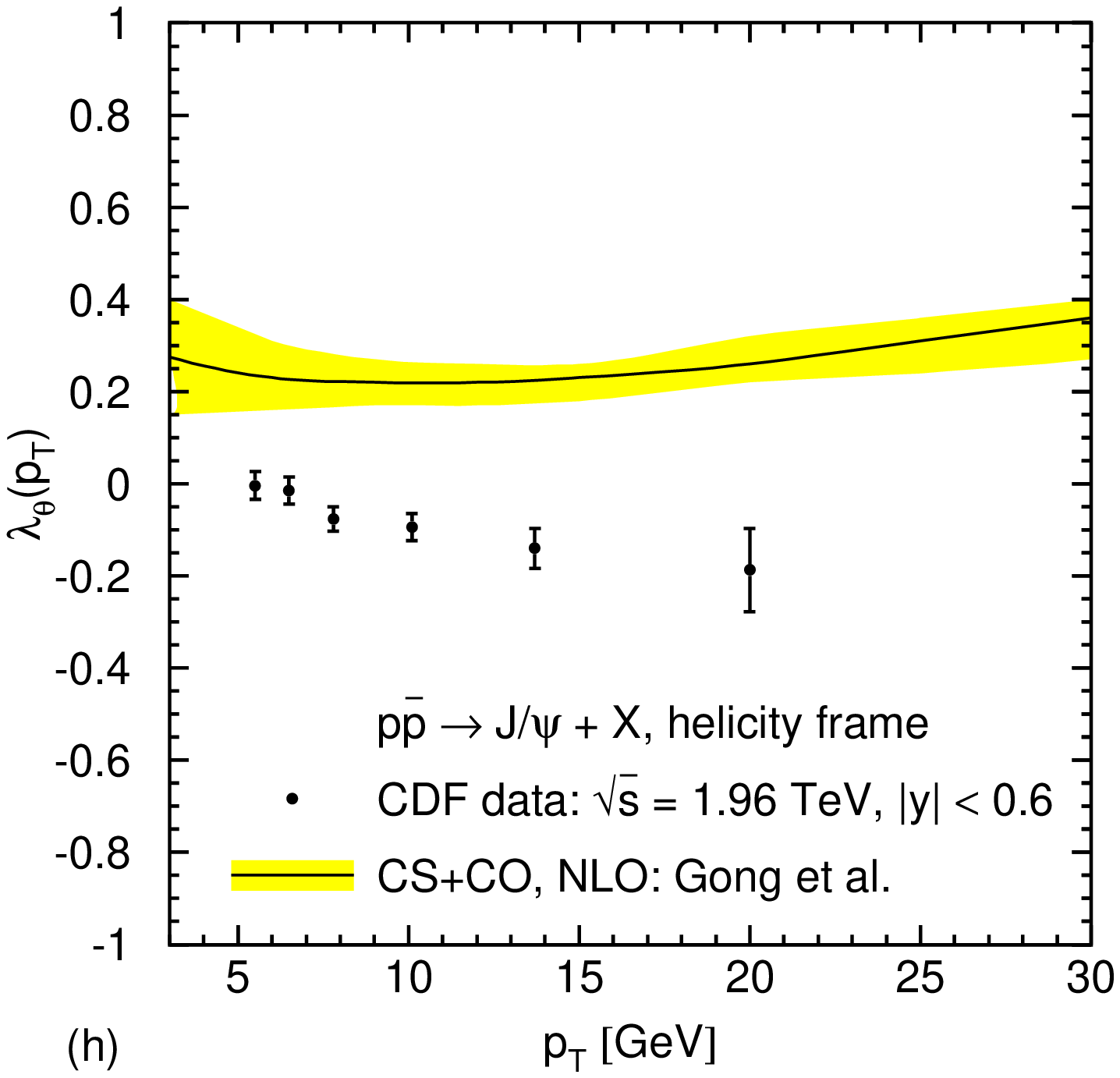}
\\
\includegraphics[width=0.22\textwidth]{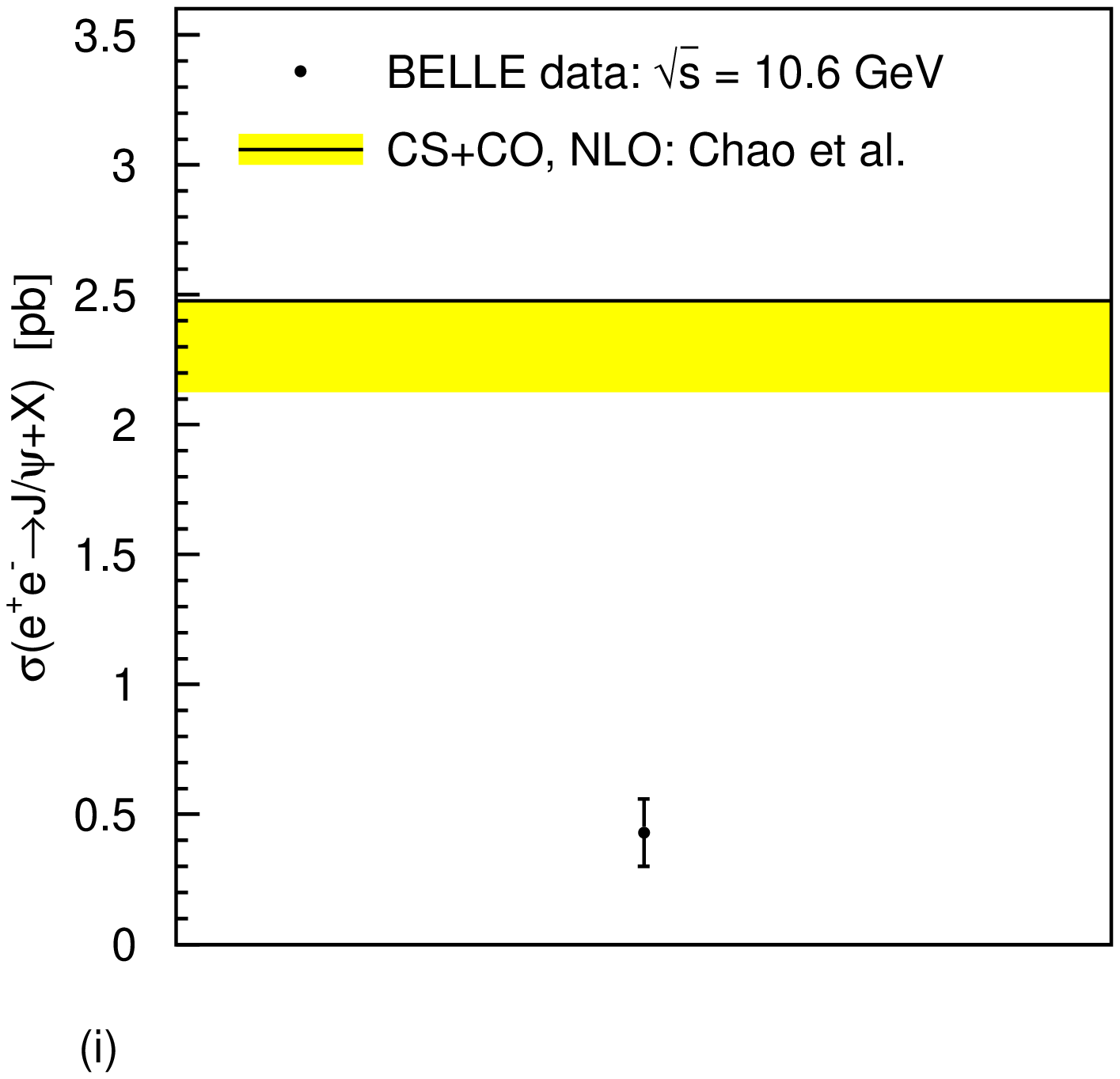}
&
\includegraphics[width=0.22\textwidth]{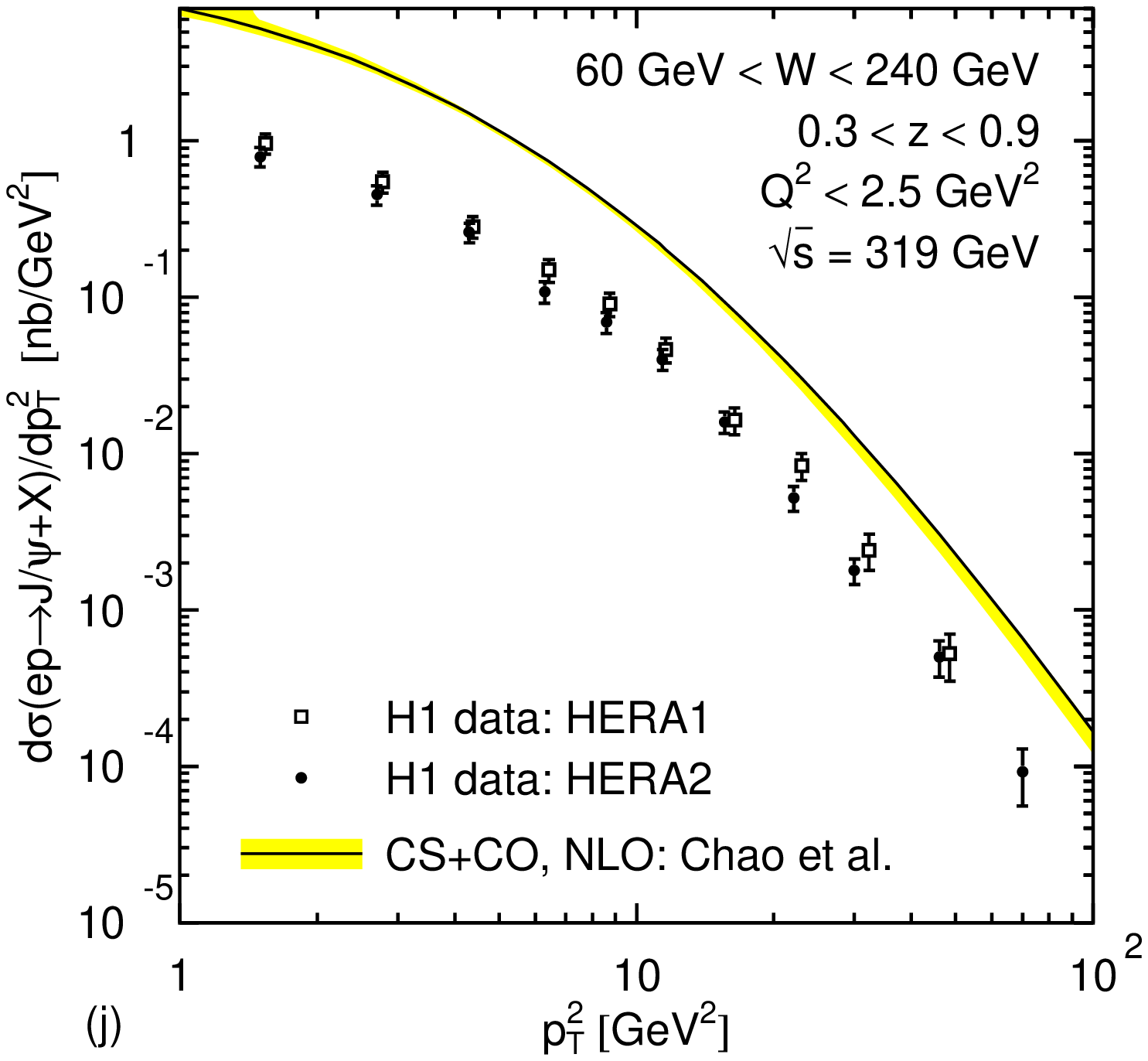}
&
\includegraphics[width=0.22\textwidth]{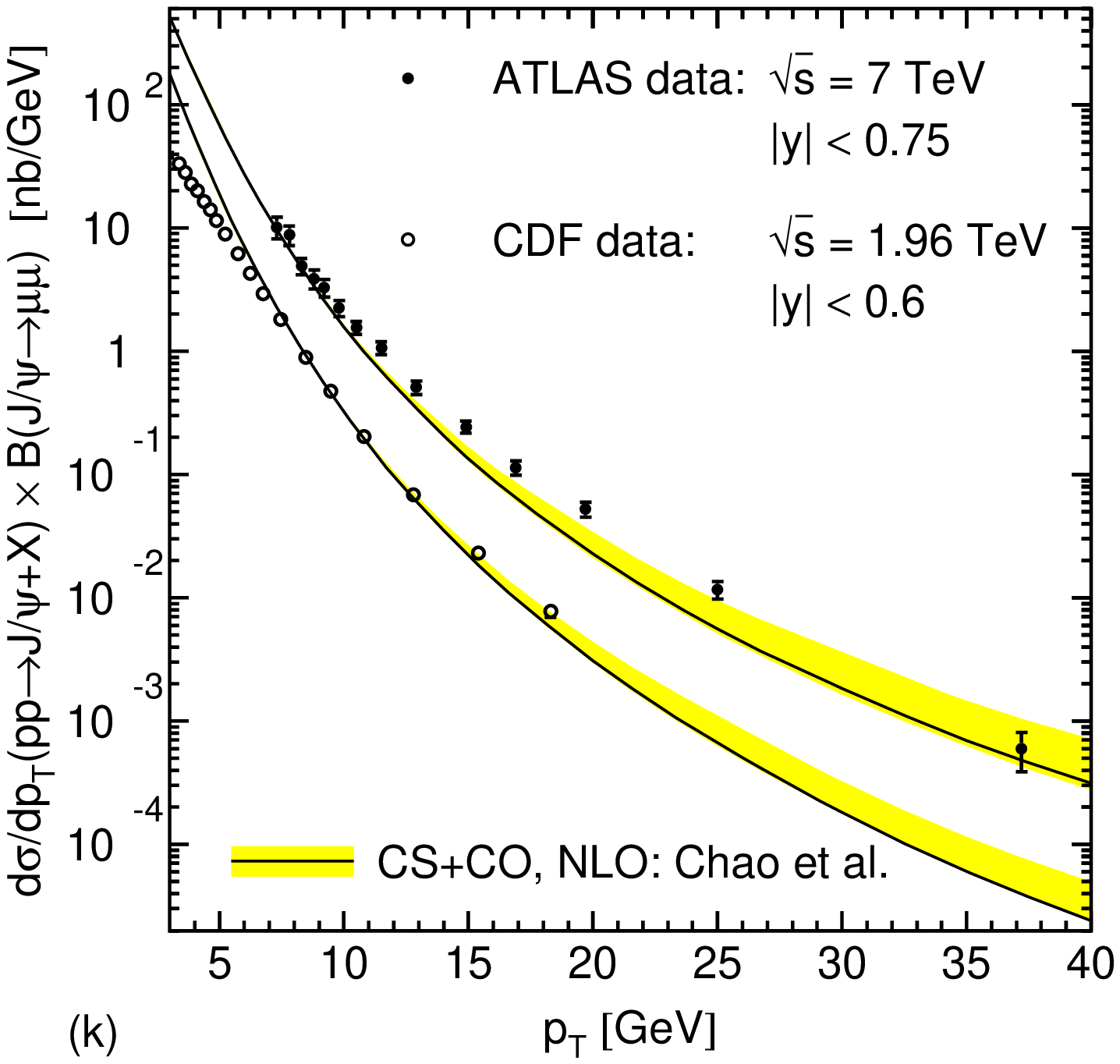}
&
\includegraphics[width=0.22\textwidth]{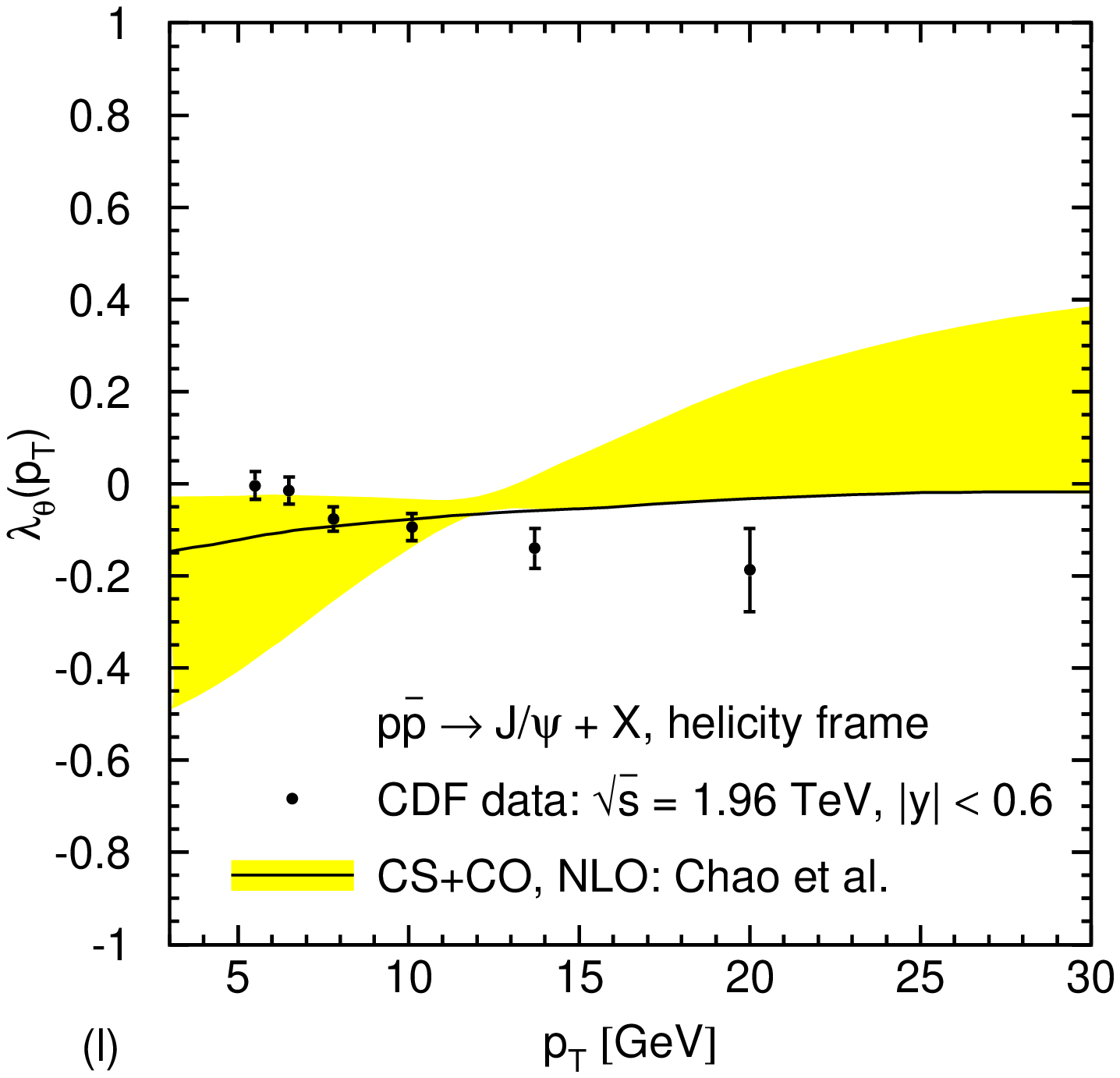}
\end{tabular}
\vspace*{8pt}
\end{center}
\caption{The unpolarized $J/\psi$ yields measured in $e^+e^-$ annihilation by
Belle,\protect\cite{Pakhlov:2009nj} in photoproduction by
H1,\protect\cite{Adloff:2002ex,Aaron:2010gz} and in hadroproduction by
CDF~II\protect\cite{Acosta:2004yw} and ATLAS\protect\cite{Aad:2011sp} as well
as the $J/\psi$ polarization observable $\lambda_\theta$ in the helicity frame
as measured by CDF~II\protect\cite{Abulencia:2007us} are compared with the NLO
NRQCD predictions evaluated using the CO LDME sets of
Refs.~\protect\refcite{Butenschoen:2011yh,Chao:2012iv,Gong:2012ug} listed in
Table~\protect\ref{tab:com}.
The theoretical errors in graphs a--g refer to scale variations, and those in
graph d are obtained by also adding in quadrature the fit errors on the CO
LDMEs according to Table~\protect\ref{tab:fit}.
Graph h is taken over from Fig.~4 of Ref.~\protect\refcite{Gong:2012ug}.
In graphs i--l, the central lines refer to the default CO LDME set of
Ref.~\protect\refcite{Chao:2012iv}, and the theoretical errors are evaluated
using the alternative CO LDME sets of Ref.~\protect\refcite{Chao:2012iv}.
\protect\label{fig:comparegraphs}}
\end{figure}

In Fig.~\ref{fig:comparegraphs}, we systematically compare the analyses of
Refs.~\refcite{Butenschoen:2012px,Butenschoen:2012qh,Chao:2012iv,Gong:2012ug}
as represented by the CO LDME sets in Table~\ref{tab:com} with regard to their
performances in describing the unpolarized $J/\psi$ yields measured in
$e^+e^-$ annihilation by Belle,\cite{Pakhlov:2009nj} in photoproduction by
H1,\cite{Adloff:2002ex,Aaron:2010gz} and in hadroproduction by
CDF~II\cite{Acosta:2004yw} and ATLAS,\cite{Aad:2011sp} as well as the $J/\psi$
polarization observable $\lambda_\theta$ in the helicity frame as measured by
CDF~II.\cite{Abulencia:2007us}
We observe that none of the LDME sets can describe all the data sets.
While the CO LDMEs of Ref.~\refcite{Butenschoen:2011yh} yield a good
description of the unpolarized $J/\psi$ yields, there is a strong disagreement
with the CDF~II measurement of $J/\psi$ polarization.
On the other hand, the CO LDMEs of Ref.~\refcite{Chao:2012iv} can describe all
hadroproduction data, but lead to overshoots by factors of 4--6 for $e^+e^-$
annihilation and photoproduction.
Finally, the CO LDMEs of Ref.~\refcite{Gong:2012ug} yield predictions which, in
all cases, fall between those of the other two options.

\begin{table}[h]
\tbl{LDME sets determined in
Refs.~\protect\refcite{Butenschoen:2011yh,Chao:2012iv,Gong:2012ug} and used in
Fig.~\protect\ref{fig:comparegraphs}.
In Ref.~\protect\refcite{Chao:2012iv}, two alternative sets are provided besides
the default one.
The analyses of
Refs.~\protect\refcite{Butenschoen:2012px,Butenschoen:2012qh,Chao:2012iv} only
refer to direct $J/\psi$ production.}
{\begin{tabular}{@{}cccccc@{}} \toprule
& Butenschoen, & Gong, Wang, & \multicolumn{3}{c}
{Chao, Ma, Shao, Wang, Zhang\cite{Chao:2012iv}} \\
& Kniehl\cite{Butenschoen:2011yh} & Wan, Zhang\cite{Gong:2012ug} &
default set & set 2 & set 3 \\
\colrule
$\langle {\cal O}^{J/\psi}(^3S_1^{[1]}) \rangle $ &
$1.32~\mbox{GeV}^3$ &
$1.16~\mbox{GeV}^3$ &
$1.16~\mbox{GeV}^3$ &
$1.16~\mbox{GeV}^3$ &
$1.16~\mbox{GeV}^3$
\\
$\langle {\cal O}^{J/\psi}(^1S_0^{[8]}) \rangle$ &
$0.0497~\mbox{GeV}^3$ &
$0.097~\mbox{GeV}^3$ &
$0.089~\mbox{GeV}^3$ &
$0$ &
$0.11~\mbox{GeV}^3$
\\
$\langle {\cal O}^{J/\psi}(^3S_1^{[8]}) \rangle$ &
$0.0022~\mbox{GeV}^3$ &
$-0.0046~\mbox{GeV}^3$ &
$0.0030~\mbox{GeV}^3$ &
$0.014~\mbox{GeV}^3$ &
$0$
\\
$\langle {\cal O}^{J/\psi}(^3P_0^{[8]}) \rangle$ &
$-0.0161~\mbox{GeV}^5$ &
$-0.0214~\mbox{GeV}^5$ &
$0.0126~\mbox{GeV}^5$ &
$0.054~\mbox{GeV}^5$ &
$0$
\\
\colrule
$\langle {\cal O}^{\psi^\prime}(^3S_1^{[1]}) \rangle$ & &
$0.758~\mbox{GeV}^3$ &
\\
$\langle {\cal O}^{\psi^\prime}(^1S_0^{[8]}) \rangle$ & &
$-0.0001~\mbox{GeV}^3$ &
\\
$\langle {\cal O}^{\psi^\prime}(^3S_1^{[8]}) \rangle$ & &
$0.0034~\mbox{GeV}^3$ &
\\
$\langle {\cal O}^{\psi^\prime}(^3P_0^{[8]}) \rangle$ & &
$0.0095~\mbox{GeV}^5$ &
\\
\hline
$\langle {\cal O}^{\chi_0}(^3P_0^{[1]}) \rangle$ & &
$0.107~\mbox{GeV}^5$ &
\\
$\langle {\cal O}^{\chi_0}(^3S_1^{[8]}) \rangle$ & &
$0.0022~\mbox{GeV}^3$ & \\ \botrule
\end{tabular}\label{tab:com}}
\end{table}

\section{Conclusions} 

As for the unpolarized $J/\psi$ yield, NRQCD factorization was consolidated at
NLO by a global fit to the world's data of hadroproduction, photoproduction,
two-photon scattering, and $e^+e^-$ annihilation,\cite{Butenschoen:2011yh}
which successfully pinned down the three CO LDMEs in compliance with the
velocity scaling rules and impressively supported their universality.
In a second step, NLO NRQCD predictions of $J/\psi$ polarization observables
in various reference frames were confronted with measurements in
photoproduction at HERA and hadroproduction at the Tevatron and the LHC.
In the case of hadroproduction at the Tevatron, the prediction of strongly
transverse $J/\psi$ polarization in the helicity frame stands in severe
contrast to the precise CDF~II measurement,\cite{Abulencia:2007us} which
found the $J/\psi$ mesons to be unpolarized.
Using the CO LDME sets recently extracted from hadroproduction data by two
other groups\cite{Chao:2012iv,Gong:2012ug} does not help us to reach a
satisfactory description of all the available precision data.
Thus, we conclude that the universality of the $J/\psi$ production LDMEs is
challenged.
Possible remedies include the following:
\begin{romanlist}[(ii)]
\item The eagerly awaited $J/\psi$ polarization measurements at the LHC might
not confirm the CDF~II results.
\item Although unlikely, measurements at a future $ep$ collider, such as the
LHeC,\cite{Armesto:2012jn} might reveal that the $p_T$ distribution of $J/\psi$
photoproduction exhibits a drastically weaker slope beyond $p_T=10$~GeV, the
reach of HERA, so that the LDME sets of Refs.~\refcite{Chao:2012iv,Gong:2012ug}
might yield better agreement with the data there.
\item The assumption that the $v$ expansion is convergent might not be valid
for charmonium, leaving the possibility that the LDME universality is intact.
\end{romanlist}

\section*{Acknowledgment}

This work was supported in part by the German Federal Ministry for Education
and Research BMBF through Grant No.\ 05H12GUE and by the Helmholtz Association
HGF through Grant No.\ Ha~101.

\end{document}